\documentclass[aps,prx,twocolumn,english,superscriptaddress,floatfix,longbibliography]{revtex4-2}

\usepackage[english]{babel}
\usepackage{amsmath}
\usepackage{units}
\usepackage{upgreek}
\usepackage{braket}
\usepackage[colorlinks=true,urlcolor=blue,linkcolor=black,citecolor=black,bookmarksopen]{hyperref}
\usepackage{orcidlink}

\newcommand*{\hatH}{\hat{H}}
\newcommand*{\hatadag}{\hat{a}^{\dagger}}
\newcommand*{\hata}{\hat{a}}
\newcommand*{\hatL}{\hat{L}}
\newcommand*{\hatLdag}{\hat{L}^{\dagger}}
\newcommand*{\finesse}{\mathcal{F}}

\begin{document}
\title{Coherent Coupling of a Diamond Tin-Vacancy Center to a Tunable Open Microcavity}
\author{Yanik Herrmann\orcidlink{0000-0003-0545-7002}}
\thanks{These authors contributed equally to this work.}
\author{Julius Fischer\orcidlink{0000-0003-2219-091X}}
\thanks{These authors contributed equally to this work.}
\author{Julia M. Brevoord\orcidlink{0000-0002-8801-9616}}
\author{Colin Sauerzapf\orcidlink{0009-0004-7655-9289}}
\altaffiliation[Present address: ]{3rd Institute of Physics and Research Center SCoPE, University of Stuttgart, 70049 Stuttgart, Germany}
\author{Leonardo G. C. Wienhoven \orcidlink{0009-0009-7745-3765}}
\author{Laurens J. Feije\orcidlink{0009-0009-6909-9145}}
\author{Matteo Pasini\orcidlink{0009-0005-1358-7896}}
\affiliation{QuTech and Kavli Institute of Nanoscience, Delft University of Technology, 2628 CJ Delft, The Netherlands}
\author{Martin Eschen\orcidlink{0009-0000-7336-6132}}
\affiliation{QuTech and Kavli Institute of Nanoscience, Delft University of Technology, 2628 CJ Delft, The Netherlands}
\affiliation{Netherlands Organisation for Applied Scientific Research (TNO), P.O. Box 155, 2600 AD Delft, The Netherlands}
\author{Maximilian Ruf\orcidlink{0000-0001-9116-6214}}
\altaffiliation[Present address: ]{SandboxAQ, Palo Alto, California, USA}
\author{Matthew J. Weaver\orcidlink{0000-0002-7093-4058}}
\altaffiliation[Present address: ]{QphoX B.V., Elektronicaweg 10, 2628 XG Delft, The Netherlands}
\author{Ronald Hanson\orcidlink{0000-0001-8938-2137}}
\email{R.Hanson@tudelft.nl}
\affiliation{QuTech and Kavli Institute of Nanoscience, Delft University of Technology, 2628 CJ Delft, The Netherlands}
\date{\today}

\begin{abstract}
Efficient coupling of optically active qubits to optical cavities is a key challenge for solid-state-based quantum optics experiments and future quantum technologies. Here we present a quantum photonic interface based on a single Tin-Vacancy center in a micrometer-thin diamond membrane coupled to a tunable open microcavity. We use the full tunability of the microcavity to selectively address individual Tin-Vacancy centers within the cavity mode volume. Purcell enhancement of the Tin-Vacancy center optical transition is evidenced both by optical excited state lifetime reduction and by optical linewidth broadening. As the emitter selectively reflects the single-photon component of the incident light, the coupled emitter-cavity system exhibits strong quantum nonlinear behavior. On resonance, we observe a transmission dip of 50 \% for low incident photon number per Purcell-reduced excited state lifetime, while the dip disappears as the emitter is saturated with higher photon number. Moreover, we demonstrate that the emitter strongly modifies the photon statistics of the transmitted light by observing photon bunching. This work establishes a versatile and tunable platform for advanced quantum optics experiments and proof-of-principle demonstrations on quantum networking with solid-state qubits.
\end{abstract}

\maketitle

\section{Introduction}
Coherent interactions between photons and two-level matter systems are a central building block in quantum optics and quantum information science, with potential application in future quantum networks for communication and computation \cite{kimble_quantum_2008,wehner_quantum_2018}. To achieve significant light-matter interaction, optical resonators (cavities) are widely employed \cite{vahala_optical_2003,reiserer_cavity-based_2015,janitz_cavity_2020,ruf_quantum_2021}. In the past years, tunable open microcavities have emerged as a versatile tool to explore and enhance the optical interface for a variety of solid-state emitters like rare-earth ions \cite{merkel_coherent_2020,casabone_dynamic_2021}, quantum dots \cite{muller_coupling_2009,herzog_pure_2018,najer_gated_2019,tomm_bright_2021}, two-dimensional materials \cite{hausler_tunable_2021, vadia_open-cavity_2021} and dye molecules embedded in organic crystals \cite{wang_coherent_2017}. These experiments capitalize on the combination of spatial flexibility, large spectral tunability, full optical accessibility, and simple hybrid integration with emitter hosts. Moreover, the mirror parameters can be tailored to a desired application, including native fiber coupling by mirrors fabricated on the tip of optical fibers \cite{hunger_fiber_2010}.\\
Much work with open microcavities has focused on diamond color centers \cite{albrecht_coupling_2013,johnson_tunable_2015,kaupp_purcell-enhanced_2016,benedikter_cavity-enhanced_2017,riedel_deterministic_2017,salz_cryogenic_2020,hoy_jensen_cavity-enhanced_2020,ruf_resonant_2021,feuchtmayr_enhanced_2023,bayer_optical_2023}, which have been the main workhorse for early quantum network nodes \cite{bernien_heralded_2013,humphreys_deterministic_2018, bhaskar_experimental_2020}. The electron spin of those color centers functions as a good communication qubit thanks to the combination of excellent quantum coherence \cite{abobeih_one-second_2018} and spin-dependent optical transitions. Furthermore, the center's native nuclear spin \cite{van_der_sar_decoherence-protected_2012,stas_robust_2022} as well as surrounding nuclear spins \cite{bradley_ten-qubit_2019} can be utilized as additional qubits, enabling network experiments with multi-qubit nodes \cite{kalb_entanglement_2017,hermans_qubit_2022,knaut_entanglement_2023}. Recent experiments on coupling color centers to open microcavities have demonstrated Purcell enhancement \cite{riedel_deterministic_2017}, resonant excitation and detection \cite{ruf_resonant_2021} and optical addressing of the spin \cite{bayer_optical_2023}. However, access to the coherent coupling regime, where the color center significantly alters the cavity transmission as well as the photon statistics of the transmitted light, has so far remained elusive.\\
In this work, we demonstrate the coherent coupling of individual diamond Tin-Vacancy (SnV) centers to a fiber-based Fabry-P\'{e}rot microcavity. We capitalize on the excellent spatial and spectral tunability of the open microcavity to select a well-coupled SnV center, which is characterized through the reduced excited state lifetime and the broadened emitter linewidth. Compared to previous implementations with Nitrogen-Vacancy centers \cite{riedel_deterministic_2017, ruf_resonant_2021}, we achieve an improvement in coherent cooperativity of nearly two orders of magnitude through the combination of SnV centers and a high quality, stable microcavity. We then exploit the coherent coupling to observe strong reduction of cavity transmission by the emitter on resonance and study the dependence on detuning and on incident photon number. Finally, we explicitly demonstrate the nonlinearity of the light-emitter interaction by measuring changes in the photon statistics of the transmitted light induced by the emitter.

\begin{figure}[ht]
    \centering
    \includegraphics[width=\linewidth]{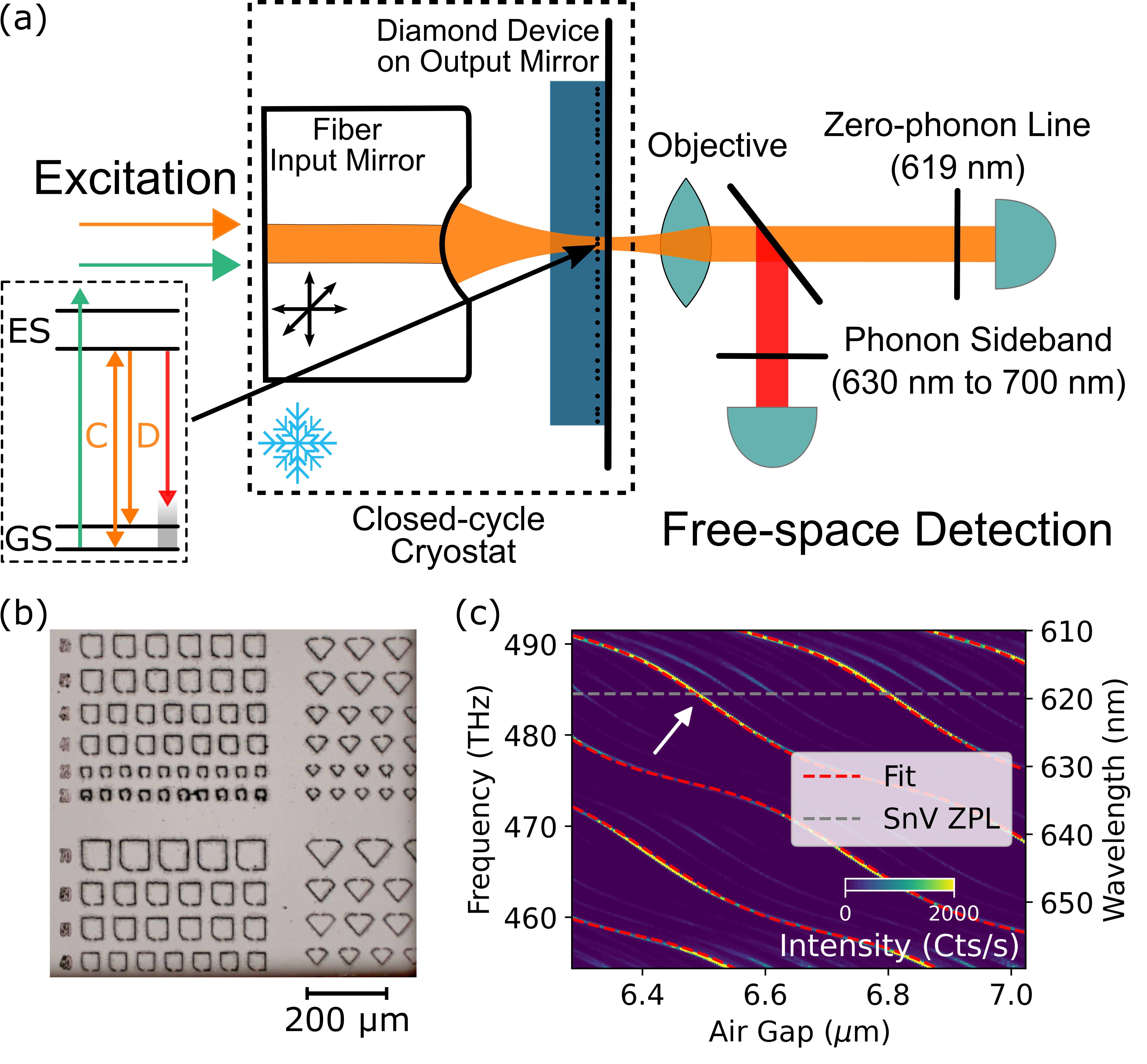}
    \caption{Schematical experimental setup, diamond devices, and air-diamond hybrid cavity operation. (a) On- and off-resonant excitation light is sent into the cavity via the fiber input mirror and the outcoupled Gaussian beam is collimated with an objective, sent to a spectrometer after free-space filtering or split into zero-phonon line and phonon sideband light and fiber-coupled to single-photon detectors (see Appendix \ref{apx:setup} for details). SnV centers are hosted in the micrometer-thin diamond devices. The inset shows the energy level structure of the negatively charged SnV center in diamond. (b) Light microscope image of the diamond devices in their holding structure. (c) Cavity transmission spectra of a broadband, white light supercontinuum source depending on the cavity air gap, which shows the mode dispersion of the air-diamond hybrid cavity. The dashed red lines display a fit using a one-dimensional hybrid cavity model \cite{janitz_fabry-perot_2015}. The white arrow indicates the operation point of the cavity at an air gap of $\unit[6.50]{\upmu m}$ and a diamond thickness of $\unit[3.72]{\upmu m}$ matching the SnV center zero-phonon line wavelength.}
    \label{fig:schematic}
\end{figure}

\section{Experimental Setup}
We use a fiber-based Fabry-P\'{e}rot microcavity, that is schematically depicted in Fig. \ref{fig:schematic} (a). The microcavity is formed by a laser-ablated concave fiber tip input mirror with a radius of curvature of $\unit[15.7]{\upmu m}$ and a macroscopic flat output mirror \cite{hunger_fiber_2010}, where thin diamond samples are bonded via van der Waals forces. The used diamond device hosts implanted SnV centers and has a thickness of about $\unit[3.72]{\upmu m}$ in the region of the cavity. SnV centers have recently emerged among solid-state color centers as a promising qubit platform with demonstrations of excellent optical and spin coherence \cite{trusheim_transform-limited_2020,rugar_narrow-linewidth_2020,arjona_martinez_photonic_2022,rosenthal_microwave_2023,guo_microwave-based_2023}. The inset of Fig. \ref{fig:schematic} (a) shows the energy level structure of the negatively charged SnV center in diamond, including the zero-phonon line (ZPL) C and D transition at $\unit[619]{nm}$ and $\unit[620]{nm}$, respectively. These transitions connect the lower branch of the excited states (ES) to the ground states (GS).\\
The sample fabrication starts with an approximately $\unit[50]{\upmu m}$ thick and top facing $\left<100\right>$ oriented diamond membrane, obtained by laser cutting and polishing \cite{ruf_optically_2019}. In the next step, $\unit[50\:\text{x}\:50]{\upmu m^2}$ squares with support bars are patterned into the diamond, followed by Tin ion implantation and subsequent annealing at $\unit[1100]{^{\circ} C}$. With the implantation energy of $\unit[350]{keV}$ (implantation dose of $\unit[3 \times 10^{10}]{ions/cm^2}$ under an angle of 7$\unit[]{^\circ}$) SnV centers in the range of the first intracavity electric field antinode located about $\unit[64]{nm}$ away from the output mirror are created. The membranes are further thinned-down to a few micrometers using a fused quartz mask and reactive ion etching \cite{appel_fabrication_2016,ruf_optically_2019}. Figure \ref{fig:schematic} (b) shows a light microscope image of the membrane after this step. Individual small diamond devices are broken out with a micromanipulator and bonded to the output mirror \cite{riedel_deterministic_2017}.\\
The input and output mirror transmittance is specified with $\unit[80]{ppm}$ and $\unit[2000]{ppm}$ (diamond termination), respectively. A piezo positioning system moves the fiber across multiple diamond devices and changes cavity position and length in situ. As illustrated in Fig. \ref{fig:schematic} (c) we operate the cavity at an air gap of $\unit[6.50]{\upmu m}$ (mode number $\unit[q = 50]{}$ of the air-diamond hybrid cavity) matching the resonance of the SnV center ZPL wavelength of $\unit[619]{nm}$ and showing an air-like mode character with a local dispersion slope of $\unit[46]{MHz/pm}$. By measuring the cavity mode in transmission with a resonant $\unit[619]{nm}$ laser a Lorentzian linewidth of $\unit[(6.86 \pm 0.05)]{GHz}$ is determined (see Appendix \ref{apx:cavity}, Fig. \ref{fig:hybrid}), resulting in a cavity quality factor $Q$ of about $7 \times 10^4$. Using a transfer matrix model we estimate an effective cavity length $L_{\text{eff}}=\unit[10.8]{\upmu m}$ and a beam waist of $\omega_0=\unit[1.24]{\upmu m}$, leading to a mode volume of $V=\unit[55]{\lambda^3}$ (see Appendix \ref{apx:cavity} for details) \cite{van_dam_optimal_2018}. The estimated total cavity losses read $\unit[7500]{ppm}$ (Finesse $\finesse = \unit[830]{}$). Comparing with the mirror transmittance values we find additional losses, that we attribute to residual scattering at the refractive index interfaces of our hybrid cavity. With these cavity parameters we calculate a Purcell factor of $F_P=\unit[6.9]{}$ following the definition
\begin{equation}
    F_P = \frac{3}{4 \pi^2} \left(\frac{\lambda}{n}\right)^3 \frac{Q}{V},
\label{equ:F_P}
\end{equation}
with the cavity resonance wavelength $\lambda$, the refractive index of diamond $n$, the cavity quality factor $Q$ and the cavity mode volume $V$.\\
The microcavity is cooled by a closed-cycle optical cryostat to a device temperature of about $\unit[8]{K}$. We measure cavity length fluctuations of $\unit[27]{pm}$ over the full cryostat cold head cycle, a five times improvement over our previous work \cite{ruf_resonant_2021} (extended technical details of the cryostat setup can be found in Ref. \cite{herrmann_low-temperature_2024}). For the SnV center measurements, excitation laser light is inserted into the cavity via the fiber input mirror. We use free-space optics to collect the light exiting the cavity through the output mirror for detection (see Fig. \ref{fig:schematic} (a)). In combination with the asymmetric mirror transmittance values, this enables the efficient coupling of the cavity mode signal into a single-mode fiber.\\
All power and photons per lifetime values stated in this work refer to the values after the output mirror of the microcavity. Moreover, the reported uncertainties correspond to one standard deviation confidence intervals.
 
\begin{figure*}%[ht]
    \centering
    \includegraphics[scale=0.8]%[width=\linewidth]
    {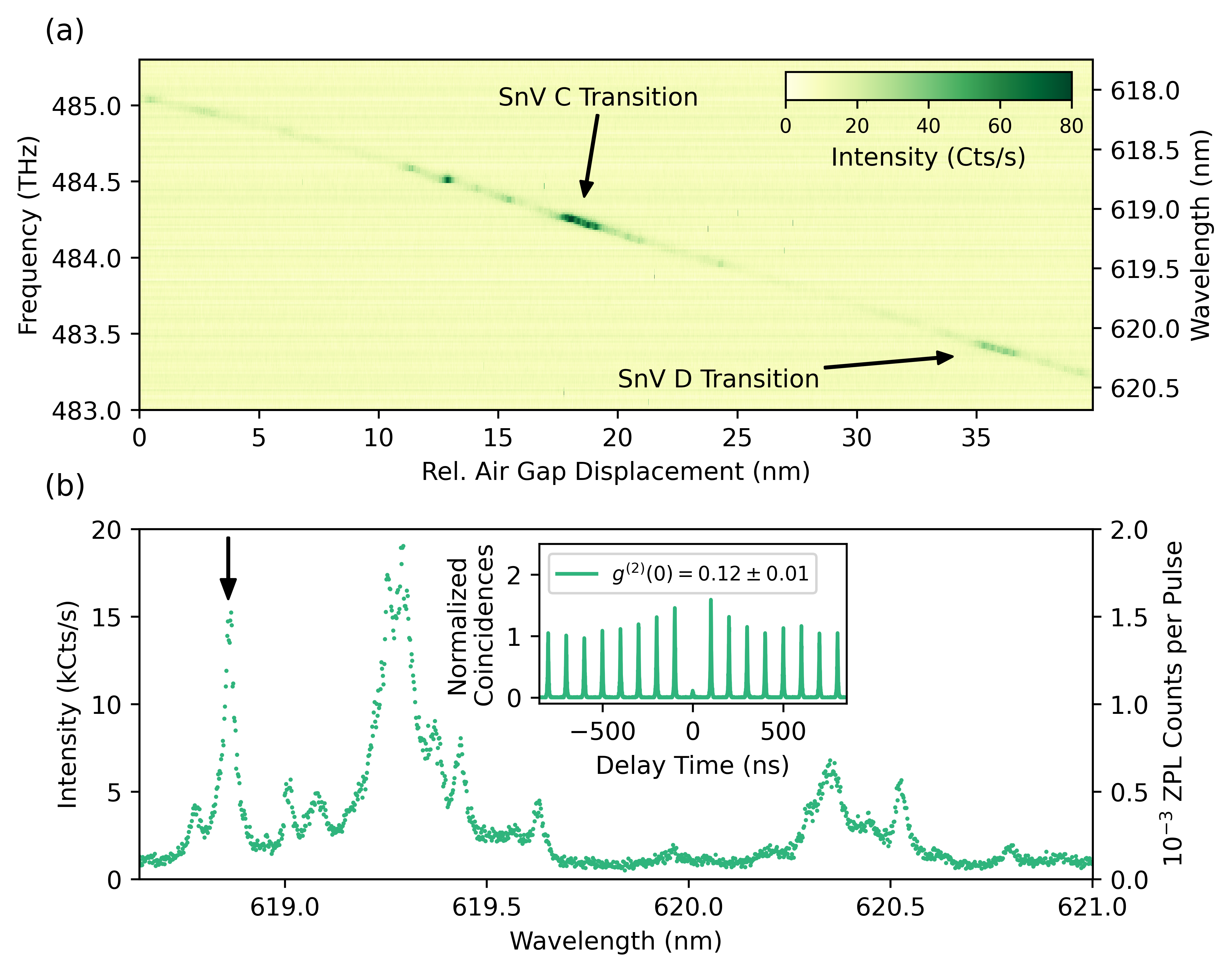}
    \caption{Coupling SnV centers to the microcavity under off-resonant excitation. (a) Analyzing the detection with a spectrometer, after filtering out the excitation light with a $\unit[600]{nm}$ longpass filter. High emission is observed when the cavity mode (straight line) becomes resonant with the C or D transition of a SnV center. (b) The detection path light is sent to a single-photon detector, after filtering with a $\unit[(620\pm5)]{nm}$ bandpass filter and polarization optics. The left and right y-axis shows the count rate and the corresponding photon detection probability per pulse (laser repetition rate $f_\text{rep}=\unit[10]{MHz}$), respectively. We tune the cavity to the isolated line, that is marked with an arrow. The inset shows a normalized $g^{(2)}(\tau)$ correlation measurement of the ZPL fluorescence, which is additionally filtered with an etalon (measurement time $\unit[20]{min}$, count rate per detector $\unit[7]{kHz}$, background count rate $\unit[200]{Hz}$).}
    \label{fig:spectrum}
\end{figure*}

\section{Coupling of individual SnV centers to the microcavity}
With the cavity resonance set close to the SnV center ZPL we explore the coupling of SnV centers in the diamond device to the optical cavity.  We excite the emitters with continuous wave, off-resonant $\unit[515]{nm}$ excitation through the input mirror and detect the light that is leaving the cavity through the output mirror. The cavity resonance frequency is tuned by applying a voltage to the piezo positioning system, on which the fiber input mirror is mounted. The cavity coating is almost transparent for $\unit[(525\pm15)]{nm}$ light, which renders the emitter excitation independent of the cavity resonance frequency. In this measurement we directly send the outcoupled light to a spectrometer after filtering out the excitation laser with a $\unit[600]{nm}$ longpass filter. When the cavity comes in resonance with an emitter, cavity-coupled SnV center light emission is expected to appear. Figure \ref{fig:spectrum} (a) shows the resulting spectra as a function of relative cavity air gap displacement. As expected for SnV centers at cryogenic temperatures, two prominent regions of high emission are observed around $\unit[619]{nm}$ and $\unit[620]{nm}$. We attribute these regions to the C and D transition of multiple SnV centers, corresponding to the optical transitions from the lower branch of the excited state to the two ground states (see energy level structure in Fig. \ref{fig:schematic} (a)).\\
To analyze the emission with the spectral resolution of the cavity (vibration averaged linewidth $\unit[8.0]{GHz}$) we scan the cavity mode over the emission lines and measure the intensity in the ZPL detection path with a single-photon detector under pulsed, off-resonant $\unit[532]{nm}$ excitation (see Fig. \ref{fig:spectrum} (b)). The ZPL detection path is equipped with a $\unit[(620 \pm 5)]{nm}$ bandpass filter and polarization optics, that are already adjusted to the emission line investigated below. In the resulting spectrum, a group of multiple SnV center emission lines around $\unit[619]{nm}$ and $\unit[620]{nm}$ are observed together with individual, spectrally isolated emission lines.\\
In the following, we focus on the isolated $\unit[619]{nm}$ emission line indicated by the arrow in Fig. \ref{fig:spectrum} (b). We additionally filter the ZPL light with an angle-tunable etalon with a full width at half maximum linewidth (FWHM) of $\approx \unit[45]{GHz}$ to further reduce background counts. This background emission may originate from parasitic light created by the green excitation laser in the cavity input fiber and in the output mirror substrate. We verify that the selected emission line corresponds to a single SnV center coupled to the cavity by measuring the $g^{(2)}(\tau)$ correlation function of the ZPL light under pulsed, off-resonant $\unit[532]{nm}$ excitation (inset of Fig. \ref{fig:spectrum} (b)). We find $g^{(2)}(0) \ll 0.5$ without background subtraction showing that indeed this emission is dominated by a single SnV center.

\begin{figure*}%[ht]
    \includegraphics[scale=0.75]{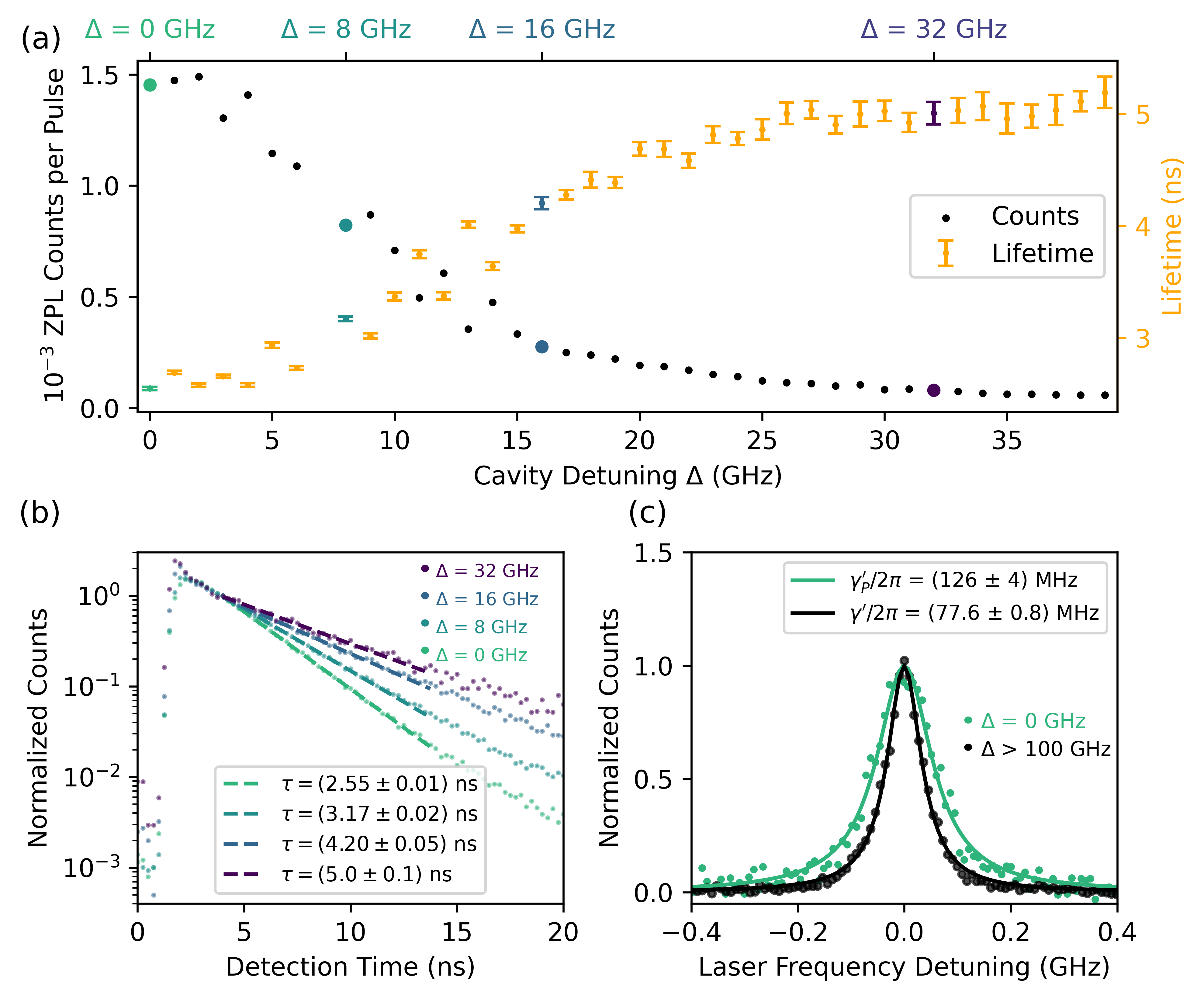}
    \caption{Quantifying the emitter-cavity coupling. (a) SnV center fluorescence and excited state lifetime for different cavity detunings with respect to the SnV center. Off-resonant, pulsed excitation ($P \approx \unit[5]{mW}$, $f_\text{rep}=\unit[10]{MHz}$) is used to excite the SnV center independently of the cavity resonance. The colored data points indicate the lifetime measurements plotted in (b). The complete dataset is shown in Appendix \ref{apx:data}, Fig. \ref{fig:full_detuning_sweep}. The error bars represent one standard deviation confidence intervals of the fit. (b) Individual lifetime measurements (integration time of $\unit[30]{s}$) of four different cavity detunings of (a), varying from fully on resonance to off resonance. The dashed lines show the fits to a monoexponential decay in a $\unit[10]{ns}$ fit window. The lifetime measurements without normalization are shown in Appendix \ref{apx:data}, Fig. \ref{fig:time_traces}. (c) On and off cavity resonance PLE linewidth measurement of the SnV center. For the off resonance measurement, the cavity is detuned by $> \unit[100]{GHz}$ with respect to the SnV center.}
    \label{fig:lifetime}
\end{figure*}

\section{Characterization of the emitter-cavity system}

After selecting an isolated SnV center emission line, we quantify the coupling of the SnV center to the cavity. We make use of the tunability of our microcavity, which enables us to switch the coupling on and off by tuning the cavity on and off resonance with the emitter. For emitters with a coupling strength to the cavity field comparable to other decay channels, the emission is significantly altered. In particular, the excited state lifetime is reduced via the Purcell effect. The coupling strength to the cavity can thus be estimated by comparing the excited state lifetimes for the cases where the cavity is on resonance and off resonance with the emitter.\\
Figure \ref{fig:lifetime} (a) shows excited state lifetime measurements of the SnV center at the C transition for different cavity frequency detunings. 
The emitter is excited by pulsed, off-resonant $\unit[532]{nm}$ excitation. Each cavity frequency detuning is set by matching the cavity resonance to a frequency-stabilized reference laser before starting the lifetime measurement. When the cavity is fully on resonance with the SnV center, we measure a detector count rate of $\unit[15]{kHz}$ and a Purcell-reduced excited state lifetime of $\tau_{P} = \unit[(2.55 \pm 0.01)]{ns}$ (see Fig. \ref{fig:lifetime} (b)). The emitter's natural lifetime of $\tau = \unit[(5.0\pm0.1)]{ns}$ (corresponding to a lifetime-limited linewidth of $\gamma = 1/2\pi\tau = \unit[32]{MHz}$) is determined at large cavity detuning, where the Purcell enhancement is negligible.\\
As expected qualitatively, the measured emitter count rate in Fig. \ref{fig:lifetime} (a) decreases with increasing cavity frequency detuning. However, the range over which the decrease occurs does not quantitatively match the previously characterized cavity linewidth (vibration averaged $\unit[8]{GHz}$). We attribute this to random emission frequency jumps (spectral diffusion) of the SnV center, due to changes in the environment caused by the strong pulsed, off-resonant $\unit[532]{nm}$ excitation light (see Appendix \ref{apx:dip} for details). The presence of this spectral diffusion in the individual lifetime measurements leads to an averaging of the Purcell enhancement. Thus, the Purcell-reduced excited state lifetime only yields a lower bound on the cooperativity, which we calculate to be $C \ge \tau/\tau_{P} - 1 = \unit[0.96 \pm 0.05]{}$.\\
We note that we observe Purcell-reduced lifetimes on several other SnV centers in this device, with the shortest lifetime measured to be $\unit[(1.78 \pm 0.01)]{ns}$ (see Appendix \ref{apx:data}, Fig. \ref{fig:large_lifetime_detung} and Fig. \ref{fig:best_snv}). Differences between measured Purcell-reduced lifetimes are likely due to differences in the cavity quality factor at different lateral positions and due to emitters being at different depths in the diamond device leading to a different overlap with the cavity field.\\
The Purcell-reduction of the excited state lifetime also reflects in a broadening of the emitter linewidth, which presents an alternative approach to measure the emitter-cavity coupling. Since spectral diffusion is expected to be strongly reduced in linewidth measurements using resonant excitation, the coupling can be more precisely determined. Figure \ref{fig:lifetime} (c) shows such photoluminescence excitation (PLE) measurements using phonon sideband (PSB) detection. We define the emitter linewidth for the case that the cavity is on resonance as $\gamma^\prime_P/2\pi$ and for a far-detuned cavity as  $\gamma^\prime/2\pi$. On cavity resonance, we measure $\gamma^\prime_P/2\pi = \unit[(126\pm4)]{MHz}$, and for a large emitter-cavity detuning of $> \unit[100]{GHz}$ an emitter linewidth of $\gamma^\prime/2\pi = \unit[(77.6\pm0.8)]{MHz}$ (see Appendix \ref{apx:PLE} for details about the PLE measurements). From these measurements we calculate a cooperativity of $C = \unit[1.7 \pm 0.2]{}$ (see Appendix \ref{apx:cooperativities} for details about the employed vibration correction) and conclude the full set of emitter-cavity parameters $\{ g,\kappa,\gamma \}/2\pi = \{ 0.30, 6.86, 0.032 \} \unit{GHz}$, with the single photon Rabi frequency $g$ and the total cavity loss rate $\kappa$. The determined cooperativity, to our knowledge the highest reported for color centers in microcavities, puts the emitter-cavity system in the coherent coupling regime in which quantum nonlinear behavior dominates the dynamics.\\
Since our cavity system is characterized and understood in detail, the measured cooperativity also allows us to draw conclusions about the SnV center properties, when considering the cooperativity definition
\begin{equation}
    C = F_P~\beta_0~\eta~\alpha~\zeta~\epsilon,
\label{equ:C_factors}
\end{equation}
with the SnV center Debye-Waller factor $\beta_0$, quantum efficiency $\eta$, branching ratio between C and D transition $\alpha$, overlap of cavity polarization with the emitter dipole $\zeta$ and the spatial overlap of cavity mode and emitter $\epsilon$. For our implementation $\zeta = \cos^2{\left(35^{\circ}\right)}$ due to the $\left<100\right>$ diamond crystal orientation. In addition, the SnV center Debye-Waller factor of $\beta_0 = 0.57 \pm 0.01$ \cite{gorlitz_spectroscopic_2020} is reported. Since the spatial overlap is bounded by unity ($\epsilon \leq 1$) a lower bound for the product of branching ratio and quantum efficiency $\alpha~\eta \ge 0.64 \pm 0.06$ is calculated. This value represents an important figure of merit to estimate the performance of future SnV-cavity implementations and is consistent with reported estimates $\eta \approx 0.8$ \cite{iwasaki_tin-vacancy_2017} and $\alpha \approx 0.8$ \cite{rugar_quantum_2021}.\\
While the cooperativity quantifies the efficiency of the spin-photon interface, the coherent cooperativity ultimately determines the fidelity of protocols \cite{borregaard_quantum_2019}. The investigated emitter exhibits contributions of nonradiative broadening above its lifetime-limited linewidth, likely due to phonon dephasing because of finite temperature effects \cite{jahnke_electronphonon_2015, wang_transform-limited_2023}. Accounting for these contributions a vibration-corrected coherent cooperativity of $C_\text{coh} = C\gamma/\gamma^\prime = \unit[0.69 \pm 0.07]{}$ is evaluated.

\begin{figure*}%[ht]
    \centering
    \includegraphics[scale=0.8]{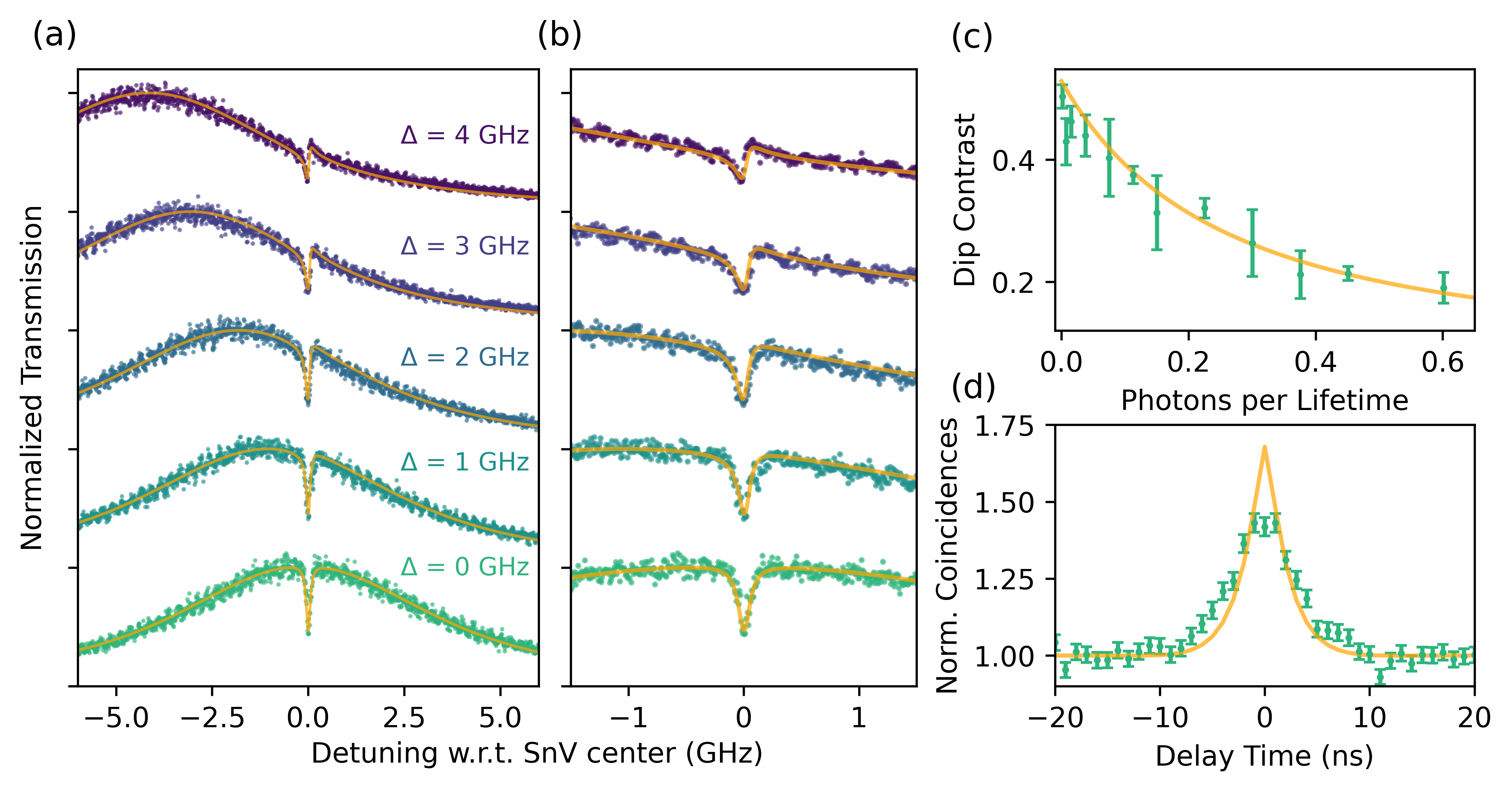}
    \caption{Nonlinear quantum effects of the emitter-cavity system. The dots are measured values, while the solid lines are modeled with a Lindblad master equation approach using the previously measured parameters of the emitter-cavity system. (a) Exemplary transmission dip measurements for increasing emitter-cavity detuning in steps of about $\unit[1]{GHz}$ and weak coherent probe laser power ($P = \unit[260]{fW}$). (b) Zoom in on the data and simulated graphs presented in (a). (c) Cavity transmission dip contrast depending on photons per Purcell-reduced excited state lifetime. The evaluation of the transmission dip contrast and error is described in Appendix \ref{apx:contrast}. (d) Measured $g^{(2)}(\tau)$ correlation function of the transmitted cavity light on emitter resonance, showing photon bunching as a quantum nonlinear effect. For the measured and simulated correlation function a binning of $\unit[1]{ns}$ is used. The error bars represent one standard deviation confidence intervals of the data.}
    \label{fig:dips}
\end{figure*}

\section{Quantum nonlinear behavior of the emitter-cavity system}
For an emitter coherently coupled to a cavity, resonant light entering the cavity is significantly modulated by coherent scattering. Due to destructive interference of the incident light with light scattered by the emitter in the forward direction, the transmission through the cavity can be strongly reduced \cite{thompson_observation_1992}. We probe this effect by scanning the frequency of a weak coherent laser and measuring the cavity transmission with a single-photon detector. Figure \ref{fig:dips} (a) depicts transmission measurements for different cavity detunings. These measurements exhibit transmission dips at the emitter frequency, with a contrast on resonance reaching $\unit[50]{\%}$. As the detuning is increased the transmission dip shape changes from absorptive to dispersive.\\
We model the emitter-cavity system with a Lindblad master equation approach \cite{sipahigil_integrated_2016}, which allows for quantitative numerical simulations of the cavity transmission as well as photon correlations (see Appendix \ref{apx:model} for details). All simulation input parameters are determined through independent measurements. The solid lines in Fig. \ref{fig:dips} (a) and (b) show the predicted, simulated transmission curves, which are in good agreement with the data and confirm our quantitative understanding of the emitter-cavity system.\\
The coherent scattering observed here is a highly nonlinear quantum effect as it results from the light interacting with a single two-level system. In Fig. \ref{fig:dips} (c) we plot the transmission dip contrasts for different light intensities, expressed in photons per Purcell-reduced excited state lifetime measured after the cavity. For the cooperativity $C=\unit[1.7]{}$ of our emitter-cavity system the Purcell-reduced lifetime reads $\unit[1.85]{ns}$. For very weak incident light intensity the emitter acts as an efficient scatterer, whereas the emitter saturates for light intensities on the order of a photon per excited state lifetime.\\
The nonlinearity of the coherent coupling fundamentally modifies the statistical properties of the transmitted light. To show this quantum nonlinear behavior we measure the $g^{(2)}(\tau)$ correlation function of the transmitted light when the cavity is on resonance with the SnV center (see Fig. \ref{fig:dips} (d)). The observed photon bunching $g^{(2)}(0)\approx 1.5$ evidence the modified statistics resulting from the selective reflection of the single-photon component of the incident weak coherent light.\\
The measured photon bunching time scale, which is related to the Purcell-reduced excited state lifetime, is slightly larger than predicted by our theoretical model (solid line). We attribute this to a residual emitter-cavity detuning during the measurement time, which leads to a slightly longer excited state lifetime (less Purcell-reduction) and thereby to a larger bunching time constant.

\section{Conclusion and Outlook}
Our work establishes a versatile tunable platform for exploring light-matter interactions with individual diamond color centers, showing coherent coupling that modifies the cavity transmission intensity and photon statistics. These results constitute the first demonstration of these quantum non-linear effects for any color center in a hybrid cavity. Furthermore, the methods used here can be directly extended to other color centers such as diamond Nitrogen-Vacancy and Silicon-Vacancy centers as well as color centers in other materials such as silicon carbide \cite{lukin_integrated_2020, heiler_spectral_2023}.\\
Our results open up near-term opportunities along several directions. First, our cavity performance may be improved by an order of magnitude as finesse values exceeding 10,000 were shown with reduced losses from the diamond sample in comparable systems \cite{bogdanovic_design_2017,hoy_jensen_cavity-enhanced_2020,flagan_diamond-confined_2022,korber_scanning_2023}. Furthermore, SnV centers in waveguides and thin membranes have shown near lifetime-limited linewidths at liquid Helium temperatures \cite{rugar_narrow-linewidth_2020,guo_microwave-based_2023}, leading to an improved coherent cooperativity.\\
Moreover, this system can be complemented with the recently established coherent control over the SnV center ground state spin \cite{rosenthal_microwave_2023,guo_microwave-based_2023}. Striplines to deliver microwaves can be fabricated onto the diamond membrane \cite{guo_microwave-based_2023} or embedded into the mirror \cite{bogdanovic_robust_2017}. Combining the light-matter interface with spin control would enable a versatile spin-photon interface for advanced quantum optic experiments and proof-of-principle demonstrations towards cavity-enhanced quantum networking with solid-state qubits.

\begin{acknowledgments}
We thank Nina Codreanu for help in the cleanroom, Henri Ervasti for software support and Lorenzo De Santis for proofreading the manuscript. We thank Johannes Borregaard, Anders S\o ndberg S\o rensen,  Robert Berghaus and Gregor Bayer for helpful discussions.\\
We acknowledge financial support from the Dutch Research Council (NWO) through the Spinoza prize 2019 (project number SPI 63-264) and from the Dutch Ministry of Economic Affairs and Climate Policy (EZK), as part of the Quantum Delta NL programme. We gratefully acknowledge that this work was partially supported by the joint research program “Modular quantum computers” by Fujitsu Limited and Delft University of Technology, co-funded by the Netherlands Enterprise Agency under project number PPS2007.\\
\textbf{Author contributions:} Y.H. and J.F. contributed equally to this work. Y.H. and J.F. conducted the experiments and analyzed the data. M.R. developed parts of the device fabrication process and designed together with M.J.W., Y.H. and J.F. the setup. Y.H., J.F., L.J.F. and M.J.W. built the setup. J.M.B., C.S. and Y.H. fabricated the diamond devices. M.E. fabricated the cavity fibers. L.G.C.W. characterized the diamond devices and the cavity fiber. J.F. and M.P. developed the simulations. Y.H., J.F. and R.H. wrote the manuscript with input from all authors. R.H. supervised the experiments.\\
\textbf{Data availability:} The datasets that support this manuscript are available at 4TU.ResearchData \cite{herrmann_data_2023}.
\end{acknowledgments}

\appendix
\section{Experimental Setup\label{apx:setup}}
\begin{figure}[ht]
    \centering
    \includegraphics[width=\linewidth]{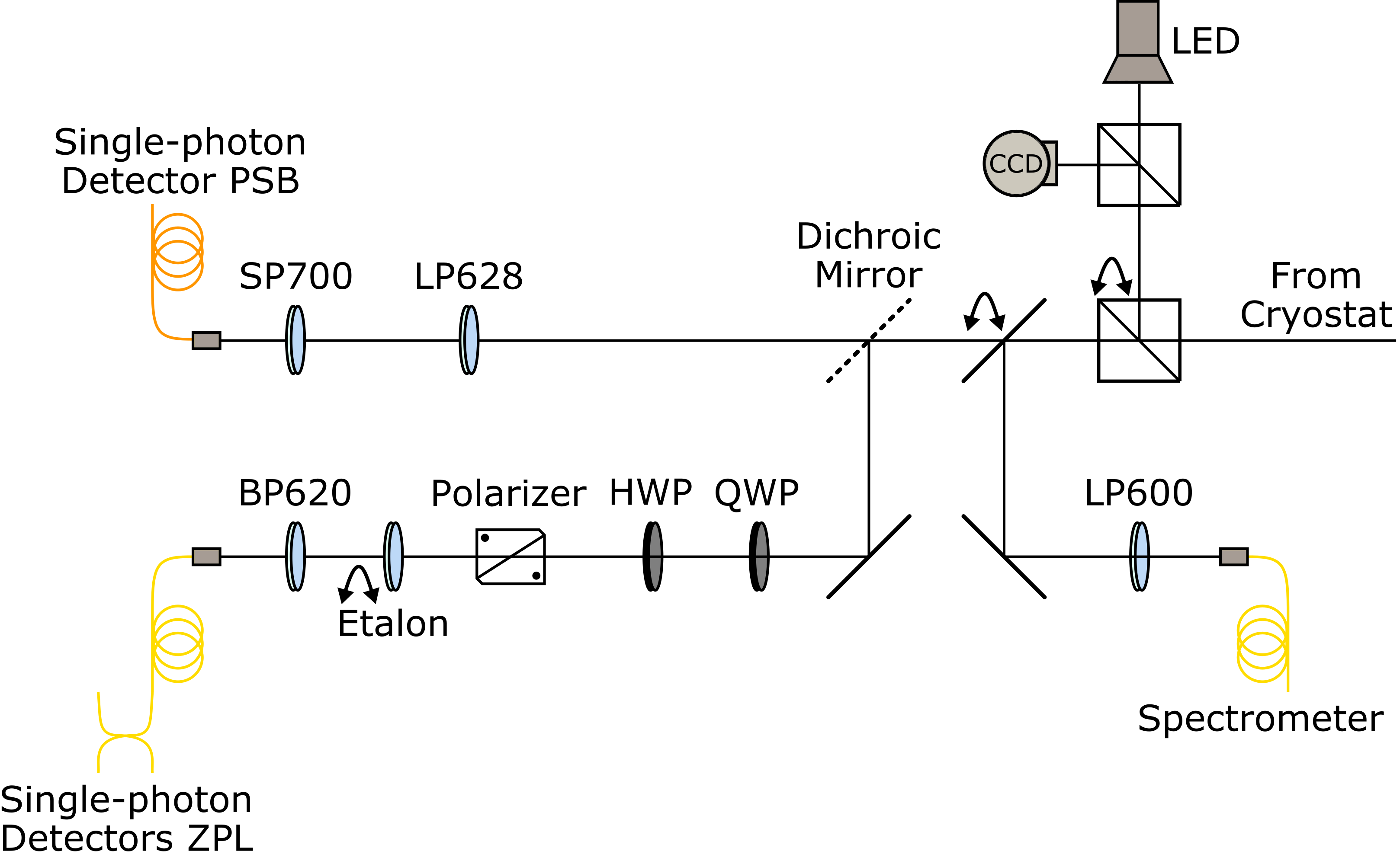}
    \caption{Schematic of the detection setup. The collimated cavity beam comes from the right-hand side and is split into ZPL and PSB detection. Alternatively, the cavity light can be sent with a flip mirror to a fiber-coupled spectrometer. Further, a 50:50 pellicle beam splitter can be inserted to image the cavity with an LED and a CCD camera. Spectral filtering is performed with various shortpass (SP), longpass (LP), and bandpass (BP) filters (see main text for details).}
    \label{fig:setup}
\end{figure}

\begin{figure}[ht]
    \centering
    \includegraphics[width=0.75\linewidth]{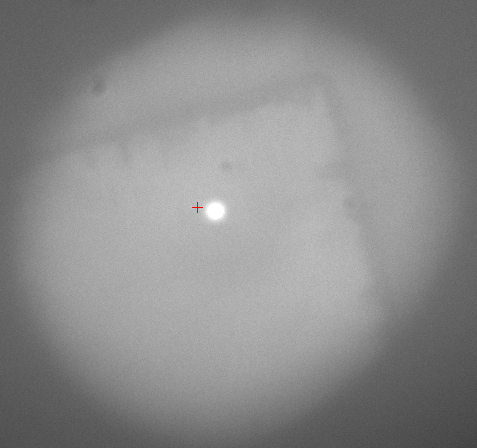}
    \caption{Micrograph of the hybrid cavity imaged through the output mirror, showing the diamond device. Off-resonant $\unit[515]{nm}$ laser light is inserted over the input mirror, illuminating the lateral cavity position used in the experiments.}
    \label{fig:sample}
\end{figure}

Experimental sequences are orchestrated with a real-time microcontroller (J\"{a}ger Adwin Pro II) and time-resolved measurements are recorded with a single photon counting module (Picoquant Hydraharp 400). The setup is controlled and measurements are performed with a PC and the Python 3 framework QMI $\unit[0.37]{}$ \cite{raa_qmi_2023}. We use quantify-core \footnote{Available under: \url{https://gitlab.com/quantify-os/quantify-core}} for data handling and analysis.\\
The excitation path consists of a frequency-doubled tunable diode laser ($\unit[619]{nm}$, Toptica TA-SHG pro), which is frequency stabilized to a wavemeter (High Finesse WS-U) and intensity-controlled with a fiber-based amplitude acousto-optic modulator (Gooch and Housego Fiber-Q 633 nm). After free-space launching, we fix the polarization with a Glan-Thompson polarizer (Thorlabs GTH10M-A), after which half- (HWP) and quarter-wave (QWP) plates are used for polarization control. For off-resonant excitation, a continuous wave laser ($\unit[515]{nm}$, H\"{u}bner Photonics Cobolt MLD515) and a pulsed laser ($\unit[532]{nm}$, $\unit[230]{ps}$ pulse duration, $\unit[10]{MHz}$ repetition rate, NKT Photonics Katana-05HP) are combined with a 30:70 beam splitter and overlapped with the $\unit[619]{nm}$ diode laser via a dichroic mirror. The light is coupled into one port of a 4x1 single mode fiber switch (Agiltron custom version), which output port is connected to the cavity fiber (single mode, $\unit[5]{\upmu m}$ mode field diameter,  IVG CU600). A different input port of the fiber switch is connected to a white light supercontinuum source (NKT Photonics SC-450-2), which is spectrally filtered for $\unit[(600-700)]{nm}$ to measure the cavity dispersion.\\
We use a floating stage Helium-free optical cryostat (Montana Instruments HILA) with an off-table cold head design and a base temperature of $\unit[6]{K}$. A detailed description of the design and operation of the full setup is presented in Ref. \cite{herrmann_low-temperature_2024}. The fiber can be placed and fine-tuned in situ with a piezo positioning stage (JPE CPSHR1-a), which is mounted on a passive vibration isolator (JPE CVIP1). The output beam of the cavity is collimated with a room temperature objective (100x magnification, $\unit[0.75]{}$ numerical aperture, $\unit[4]{mm}$ working distance, Zeiss LD EC Epiplan-Neofluar), which is positioned by three linear piezo stages (Physik Instrumente Q-545) in a tripod configuration. \\
The collimated cavity beam leaves free-space the cryostat vacuum chamber to the detection path shown in Fig. \ref{fig:setup}. We illuminate and monitor the cavity with an LED lamp and a CCD camera to position the fiber with respect to the sample (see Fig. \ref{fig:sample} for the exact cavity spot on the diamond device). To measure the cavity dispersion or SnV center fluorescence, the light can be sent to a fiber-coupled spectrometer (Princeton Instruments SP-2500i). For all other measurements, the light is split into the ZPL and PSB path with a longpass filter (Asahi Spectra LP $\unit[630]{nm}$). The PSB path is spectrally filtered with two longpass (Semrock VersaChrome Edge TLP01-628) and a shortpass filter (Thorlabs FES700) and fiber-coupled into a multimode fiber ($\unit[25]{\upmu m}$ core size, $\unit[0.1]{}$ numerical aperture, Thorlabs FG025LJA) with an objective (10x magnification, $\unit[0.1]{}$ numerical aperture, $\unit[10.6]{mm}$ working distance, Olympus RMS10X). The PSB light is measured with a single-photon detector (Laser Components COUNT-10C-FC). The ZPL light is filtered for polarization with a quarter- and half-wave plate and a Glan-Thompson polarizer (Thorlabs GTH10M-A) and spectrally with a bandpass filter (Thorlabs FBH620-10). We additionally filter the light for some experiments with an angle-tunable free-space etalon (full width at half maximum $\approx\unit[45]{GHz}$, free spectral range $\approx\unit[2.7]{THz}$, LightMachinery custom coating). The ZPL light is fiber-coupled into a single mode fiber ($\unit[3.6-5.3]{\upmu m}$ mode field diameter, $\unit[0.1-0.14]{}$ numerical aperture, Thorlabs SM600) with an objective (Olympus RMS10X). For the correlation measurements, a 50:50 single-mode fiber beam splitter (Thorlabs TW630R5A2) is used. The ZPL light is measured with a single-photon detector (Picoquant Tau-SPAD-20).

\section{Hybrid Cavity Characterization and Simulation\label{apx:cavity}}
\begin{figure}[ht]
    \centering
    \includegraphics[width=\linewidth]{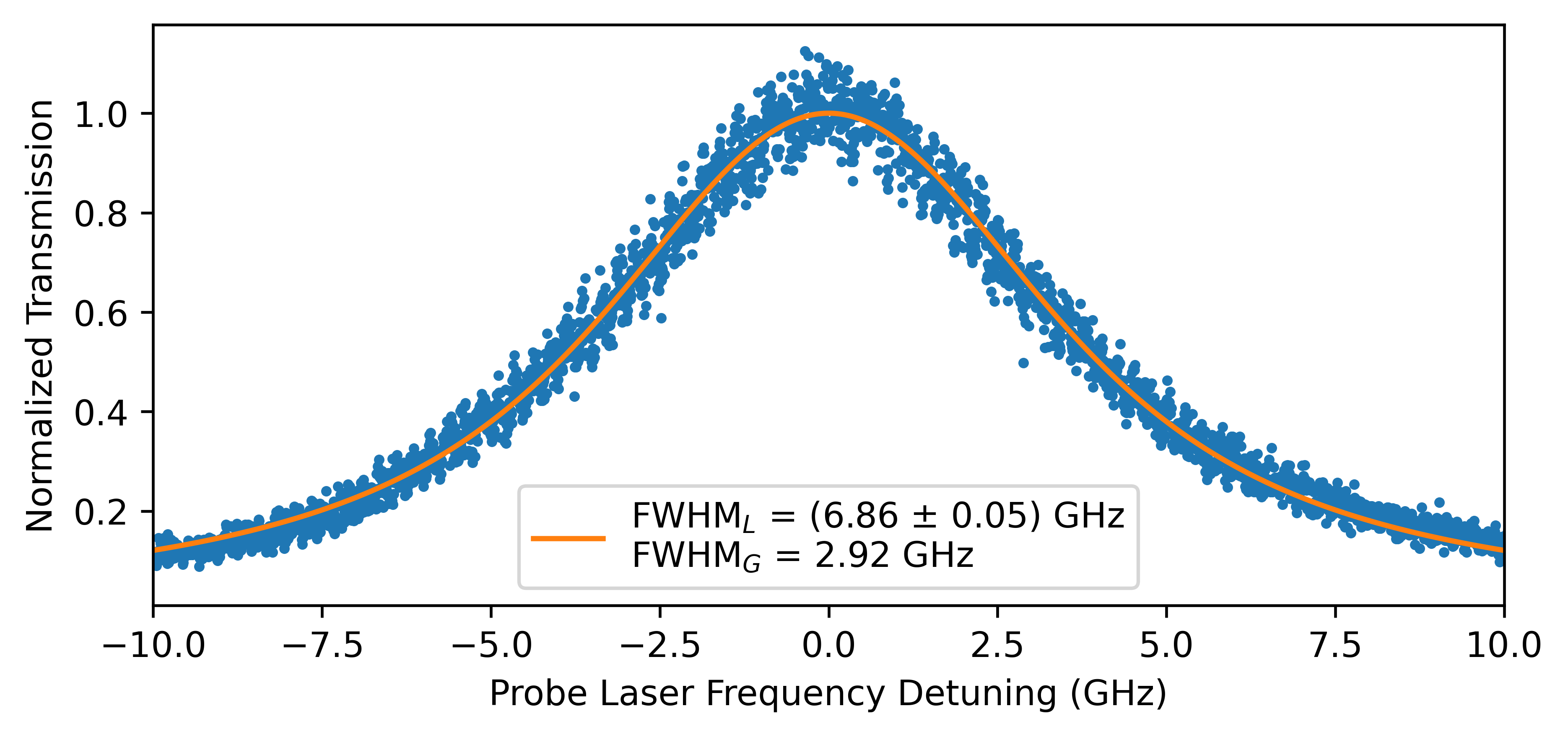}
    \caption{Linewidth characterization of the air-diamond hybrid cavity. Extracted from a resonant cavity transmission scan fit, the cavity shows a total Voigt linewidth of $\unit[8]{GHz}$, composed of a Lorentzian cavity linewidth of $\unit[(6.86\pm0.05)]{GHz}$ and a Gaussian contribution of $\unit[2.92]{GHz}$ ($=\unit[27]{pm} \cdot \unit[46]{MHz/pm} \cdot 2\sqrt{2\ln{2}}$) accounting for $\unit[27]{pm}$ RMS cavity length fluctuations.}
    \label{fig:hybrid}
\end{figure}

Figure \ref{fig:hybrid} shows a cavity transmission measurement used to determine the cavity linewidth. In this measurement, a resonant laser is scanned over the cavity resonance, while the transmission is detected with a single-photon detector in the ZPL path. To this transmission signal a Voigt fit with a fixed Gaussian contribution of $\unit[2.92]{GHz}$ is applied. This Gaussian contribution accounts for the $\unit[27]{pm}$ root mean square (RMS) value of cavity length fluctuation. The Voigt fit yields a Lorentzian contribution of $\unit[(6.86\pm0.05)]{GHz}$, which corresponds to the Lorentzian cavity linewidth.

\begin{figure}[ht]
    \centering
    \includegraphics[width=\linewidth]{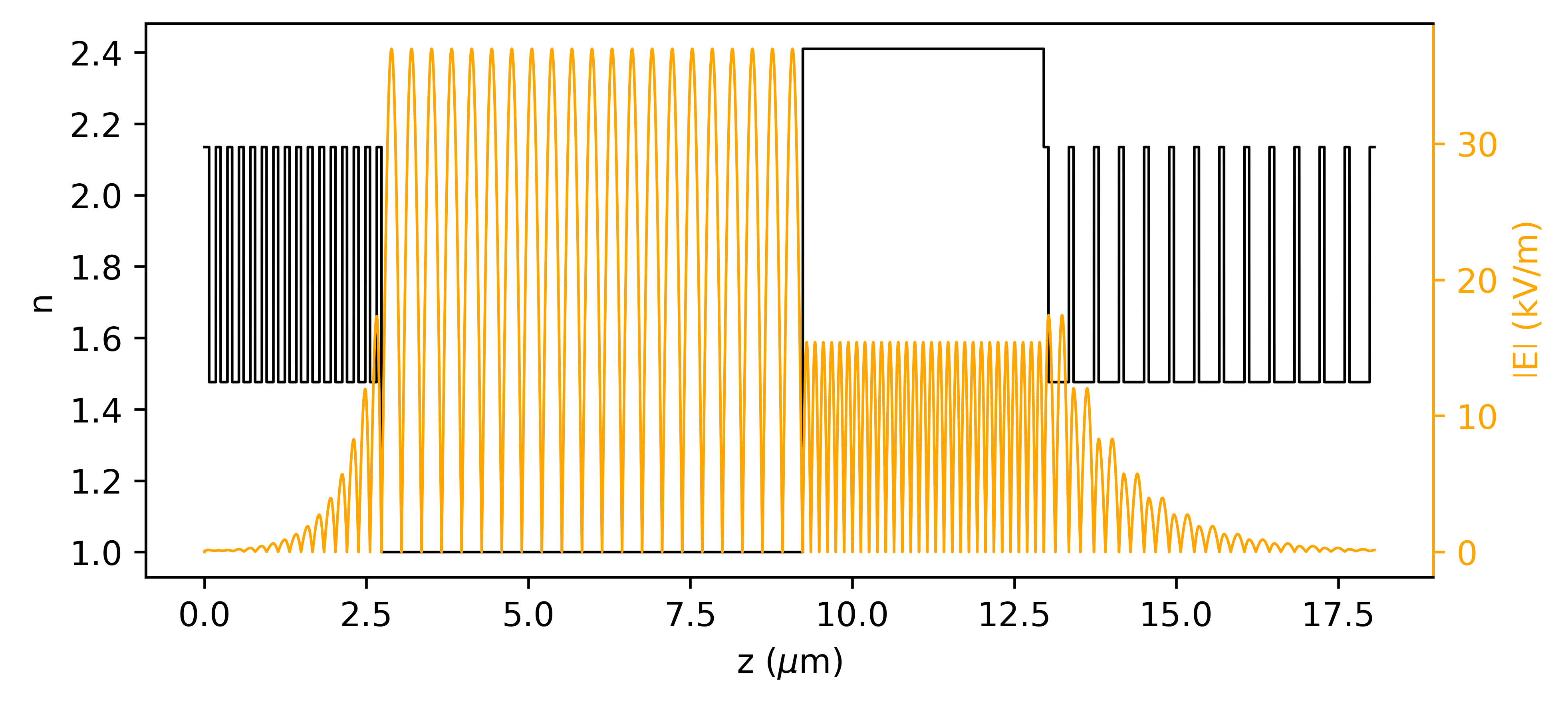}
    \caption{Simulation of the electric field distribution inside the microcavity. The solid black line displays the refractive index $n$ over the extent of our microcavity and the yellow solid line shows the simulated electric field strength $\left|E\right|$. The dielectric input and output mirror are shown on the left and right side. In between, the air gap of $\unit[6.50]{\upmu m}$ and the diamond of thickness $\unit[3.72]{\upmu m}$ are modeled with a refractive index of $n=1$ and $n=2.41$, respectively. An effective cavity length of $L_\text{eff} = \unit[10.8]{\upmu m}$ is numerically calculated.}
    \label{fig:cavity_field}
\end{figure}

To calculate the cavity mode volume of our microcavity the electric field distribution inside the cavity is simulated in Fig. \ref{fig:cavity_field} \cite{van_dam_optimal_2018}. The electric field is confined between the dielectric input and output mirror and spreads over the air and diamond part in between. From simulations an effective cavity length $L_\text{eff} = \unit[10.8]{\upmu m}$ is calculated, which leads with the cavity beam waist $\omega_0 = \unit[1.24]{\upmu m}$ to a cavity mode volume of $\unit[55]{\lambda^3}$.

\section{Vibration Model\label{apx:vib}}
The cavity length fluctuations of our microcavity lead to vibration averaging in measurements, which we take into account in simulations and when stating vibration-corrected quantities. We model a Gaussian distribution of cavity length fluctuations, that translate over a local linear cavity mode dispersion into cavity frequency detunings $\Delta \nu$. The probability density function is given by \cite{ruf_resonant_2021}
\begin{equation}
    f(\Delta\nu) = \frac{1}{\sqrt{2\pi\sigma^2}} e^{-\Delta\nu^2/2\sigma^2},
\label{equ:Gaussian_prob}
\end{equation}
with the RMS value $\sigma$ of the cavity frequency fluctuations. We perform the vibration averaging by discretizing the probability density function and integrating numerically.

\section{Cooperativity Definitions and Vibration Correction\label{apx:cooperativities}}
The efficiency of a spin-photon interface is quantified by the cooperativity \cite{borregaard_quantum_2019}
\begin{equation}
    C = \frac{4g^2}{\kappa\left(\gamma_\text{rad} + \gamma_\text{nonrad}\right)},
\label{equ:C}
\end{equation}
with the single photon Rabi frequency $g$ and the emitter decay rate $\gamma = \gamma_\text{rad} + \gamma_\text{nonrad}$ with its radiative and nonradiative component. For the SnV center the radiative component is composed of a ZPL and a PSB part $\gamma_\text{rad} = \gamma_\text{ZPL} + \gamma_\text{PSB}$. Further, the ZPL component splits up into the SnV center C and D transition $\gamma_\text{ZPL} = \gamma_\text{C} + \gamma_\text{D}$.

In this study, we couple a narrow-linewidth SnV center via its C transition to a spectrally broad cavity ($\kappa\gg\gamma$). The cavity-induced increase in the C transition decay rate with respect to its initial value $\gamma_\text{C}$ is determined by the Purcell factor times the spatial $\epsilon$ as well as polarization $\zeta$ overlap between emitter and cavity mode. Hence, the Purcell-enhanced emitter decay rate reads
\begin{equation}
    \gamma_P = F_P~\epsilon~\zeta~\gamma_\text{C} + \gamma.
\label{equ:gamma_P1}
\end{equation}
This Purcell-enhanced decay rate can be expressed as a function of the cooperativity by introducing the branching ratio between C and D transition $\alpha = \gamma_\text{C}/\gamma_\text{ZPL}$, the Debye-Waller factor $\beta_0 = \gamma_\text{ZPL}/\gamma_\text{rad}$ and the quantum efficiency $\eta=\gamma_\text{rad}/\gamma$. Equation \eqref{equ:gamma_P1} transforms to
\begin{equation}
    \gamma_P = F_P~\epsilon~\zeta~\alpha~\beta_0~\eta~\gamma + \gamma = \gamma (C + 1),
\label{equ:gamma_P2}
\end{equation}
with the cooperativity $C$ as defined in equation \eqref{equ:C_factors} of the main text.\\
Following equation \eqref{equ:gamma_P2} the cooperativity
\begin{equation}
    C = (\gamma_P-\gamma)/\gamma = \tau/\tau_P - 1,
\label{equ:C_lifetime}
\end{equation}
can be determined by measuring the Purcell-reduced excited state lifetime $\tau_P$ and the natural excited state lifetime $\tau$. In the case of a linewidth-broadened emitter $\gamma^\prime = \gamma + \gamma_\text{dp}$ with pure dephasing rate $\gamma_\text{dp}$ the cooperativity
\begin{equation}
    C = (\gamma^\prime_P - \gamma^\prime)/\gamma,
\label{equ:C_lifetime_dephasing}
\end{equation}
can be calculated by measuring the emitter linewidth $\gamma^\prime$ and the Purcell-enhanced linewidth $\gamma^\prime_P$ together with the lifetime-limited linewidth $\gamma = 1/2\pi\tau$ deduced from a lifetime measurement.\\
The coherent cooperativity that takes emitter coherence into account reads \cite{borregaard_quantum_2019}
\begin{equation}
    C_\text{coh} = C \frac{\gamma}{\gamma^\prime} = \frac{\gamma^\prime_P}{\gamma^\prime} - 1,
\label{equ:C_coh}
\end{equation}
and hence can be determined by measuring the emitter linewidth and the Purcell-enhanced linewidth only.\\
Moreover, the emitter-cavity coupling, described by the cooperativity $C$, depends on the spectral overlap between emitter and cavity. In the regime where the emitter linewidth is much smaller than the cavity linewidth the cooperativity is given by \cite{van_dam_optimal_2018}
\begin{equation}
    C_\text{overlap}(\nu_{e},\nu_{c}) = \frac{C}{1 + 4 Q^2 (\nu_{e}/\nu_{c}-1)^2},
\label{equ:Spectral_overlap}
\end{equation}
with the maximal cooperativity $C$ on emitter-cavity resonance, the cavity quality factor $Q$, the cavity resonance frequency $\nu_{c}$ and the emitter transition frequency $\nu_{e}$. Note that this dependency needs generally to be considered for emitter-cavity detunings and also vibration averaging. Using equation \eqref{equ:Spectral_overlap} and \eqref{equ:Gaussian_prob} a ratio of $\unit[0.90]{}$ between measured (vibration averaged) and maximal cooperativity on emitter-cavity resonance is calculated for our RMS vibration level of $\unit[27]{pm}$. We take this factor into account when stating vibration-corrected cooperativity values.

\section{Modeling SnV Centers in Optical Cavities\label{apx:model}}
The investigated emitter-cavity system under resonant excitation is described by an atomic system and a driven optical cavity, modeling the SnV center and the probed microcavity, respectively. For sample temperatures of about $\unit[8]{K}$ the SnV center in diamond is well approximated by a two-level atomic system, considering only the lower branches of the excited and ground states that are linked via the SnV center C transition.\\
The population in the upper branch states of the ground state can be calculated by the ratio of the fast electron-phonon transition rates between the ground state branches $\gamma_+$ and $\gamma_-$ \cite{jahnke_electronphonon_2015}. The ratio $\gamma_+/\gamma_-$ is determined by the thermal population of the corresponding phonon mode, which for the SnV center ground state splitting of $\approx \unit[850]{GHz}$ and temperature of $\unit[8]{K}$ is well approximated by the Maxwell-Boltzmann distribution. In this regime the population of the upper branch states follows the Maxwell-Boltzmann distribution as well and is with $< \unit[1]{\%}$ very low, allowing to disregard this state. Note that the SnV center Debye-Waller factor, quantum efficiency, and branching ratio are included in the cooperativity of the emitter-cavity system.\\
The dynamics of the emitter-cavity system are modeled with a Lindblad master equation approach, where the Hamiltonian of the system in the rotating frame of the probe frequency reads \cite{sipahigil_integrated_2016}
\begin{equation}
\begin{split}
    \hatH = &\Delta_e \ket{e}\bra{e} + \Delta_c \hatadag \hata + i \xi \left( \hatadag - \hata \right) \\
     &+ i g \left( \hata \ket{e}\bra{g} - \hatadag \ket{g}\bra{e} \right)
\end{split}
\label{equ:Hamiltonian}
\end{equation}
with $\Delta_e/2\pi = \nu_{e} - \nu_{p}$, $\Delta_c/2\pi = \nu_c - \nu_p$, where $\nu_{e}$ is the emitter transition frequency between excited state $\ket{e}$ and ground state $\ket{g}$ and $\nu_c$ the resonance frequency of the cavity. The cavity field is described by the bosonic photon annihilation operator $\hata$. The single photon Rabi frequency $g$ determines the coupling strength between the two atomic levels and the cavity field via the Jaynes-Cummings interaction. The weak coherent probe laser field at frequency $\nu_p$ has a photon flux amplitude of $\xi / \sqrt{\kappa_{in}}$ at the cavity input mirror. The Lindblad operators of the system are
\begin{align}
    \hatL_1 &= \sqrt{\gamma} \ket{g}\bra{e},\nonumber\\
    \hatL_2 &= \sqrt{\gamma_\text{dp}} \ket{e}\bra{e},\nonumber\\
    \hatL_3 &= \sqrt{\kappa} \hata,
\label{equ:Linblad_operators}
\end{align}
with the decay rate $\gamma$ from the state $\ket{e}$ to $\ket{g}$, the dephasing rate $\gamma_\text{dp}$ and the total cavity loss rate $\kappa$.
The Linblad master equation for the system density matrix $\rho$ reads
\begin{equation}
    \dot{\rho} = - \left[ \hatH, \rho \right] + \sum_{x=1}^{3} \hatL_x \rho \hatLdag_x - \frac{1}{2} \left( \hatLdag_x \hatL_x \rho + \rho \hatLdag_x \hatL_x \right).
\label{equ:Master_Equation}
\end{equation}
In the weak driving regime ($\xi \ll \kappa$) the Hilbert space can be truncated and we assume at most two excitations in our system, which results in the basis
\begin{equation}
    \left\{ \ket{0,g},\ket{0,e},\ket{1,g},\ket{1,e},\ket{2,g} \right\}.
\label{equ:Basis}
\end{equation}
To calculate the cavity transmission and intensity correlation function of the transmitted light the steady state of the system is considered. The cavity transmission is given by
\begin{equation}
    T = \frac{\kappa_{out} \braket{\hatadag \hata}}{\braket{\hatadag_{in} \hata_{in}}} = \frac{\kappa_{out} \kappa_{in}}{\xi^2} \braket{\hatadag \hata},
\label{equ:Transmission}
\end{equation}
with the loss rate of the output mirror $\kappa_{out}$ and the input photon flux $\braket{\hatadag_{in} \hata_{in}} = \xi^2/\kappa_{in}$.
The normalized intensity correlation function of the transmitted light is given by
\begin{equation}
    g^{(2)}_T(\tau) = \frac{\braket{\hatadag(0)\hatadag(\tau) \hata(\tau)\hata(0)}}{\braket{\hatadag \hata}^2},
\label{equ:G2}
\end{equation}
with the correlation time delay $\tau$.\\
We use the Python toolbox qutip \cite{johansson_qutip_2012} to solve for the operator expectation values. To calculate the photon field expectation values of the correlation functions we increase the photon Hilbert space to eight excitations for numerical stability.

\section{Further Data on SnV-Cavity Coupling\label{apx:data}}
\begin{figure}[ht]
    \centering
    \includegraphics[width=\linewidth]{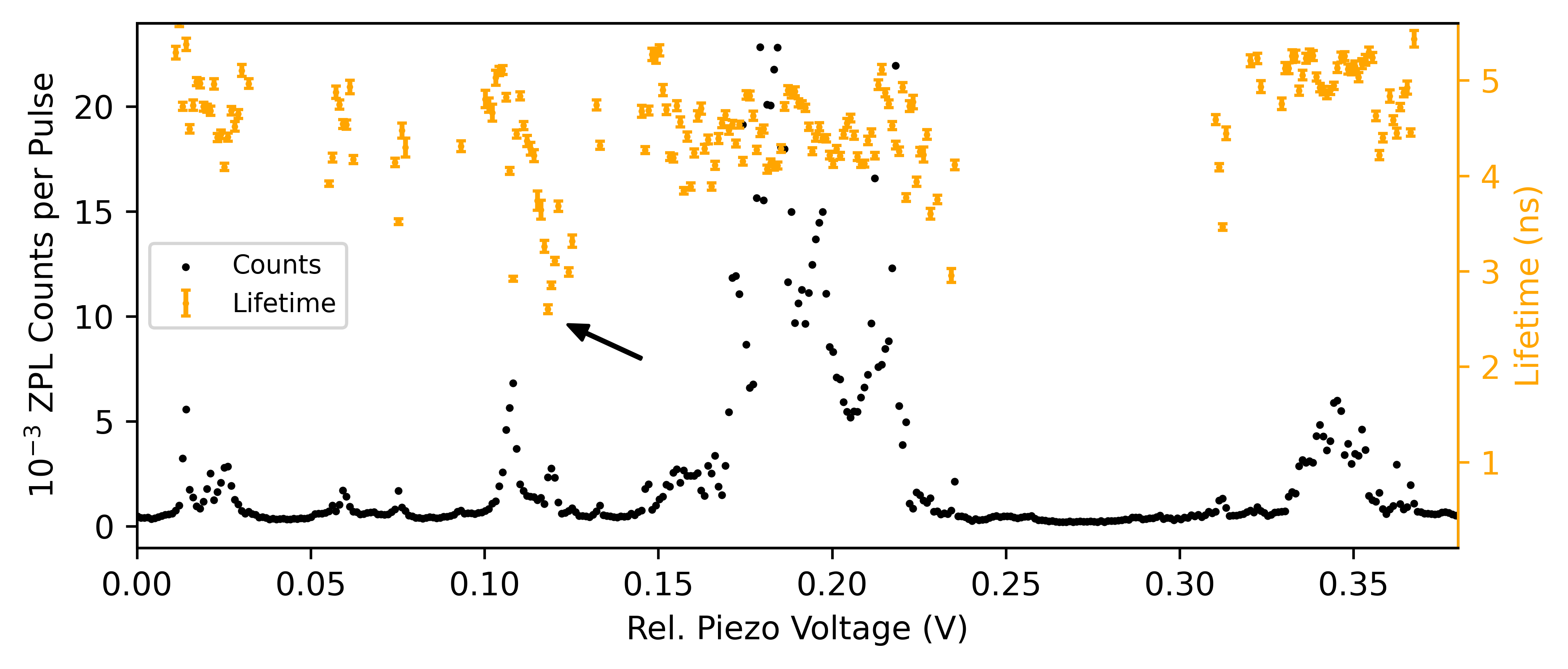}
    \caption{Detuning of the cavity resonance from about $\unit[617]{nm}$ to $\unit[621]{nm}$ with a lifetime measurement triggered on a predefined count rate threshold of $\unit[7]{kCts/s}$ (background of $\unit[2]{kCts/s}$). The ZPL path is filtered with a $\unit[(620\pm5)]{nm}$ bandpass filter only. We attribute the clustered peaks around $\unit[0.2]{V}$ and $\unit[0.35]{V}$ to the C and D transition of multiple SnV centers. The SnV center used in this study is indicated by the black arrow with a Purcell-reduced excited state lifetime of about $\unit[2.5]{ns}$. The error bars represent one standard deviation confidence intervals of the fit.}
    \label{fig:large_lifetime_detung}
\end{figure}

To identify SnV centers exhibiting large cavity coupling strengths within the cavity mode volume, we tune the cavity resonance over a large range by sweeping the piezo voltage of the fiber positioning system. We simultaneously excite with a pulsed, off-resonant $\unit[532]{nm}$ laser and monitor the counts in the ZPL detection. Once a certain count rate is exceeded, the sweep is paused and a lifetime measurement (usually $\unit[10]{s}$ to $\unit[60]{s}$ integration time) is performed to extract the excited state lifetime. The measurement and analysis is fully automatized and the best coupled SnV centers are found efficiently. For the cavity used in this work, an exemplary sweep is shown in Fig. \ref{fig:large_lifetime_detung}, where the SnV center used in this study is located at a piezo voltage of $\unit[0.12]{V}$ with a Purcell-reduced excited state lifetime of about $\unit[2.5]{ns}$ as indicated by the black arrow.

\begin{figure}[ht]
    \centering
    \includegraphics[width=\linewidth]{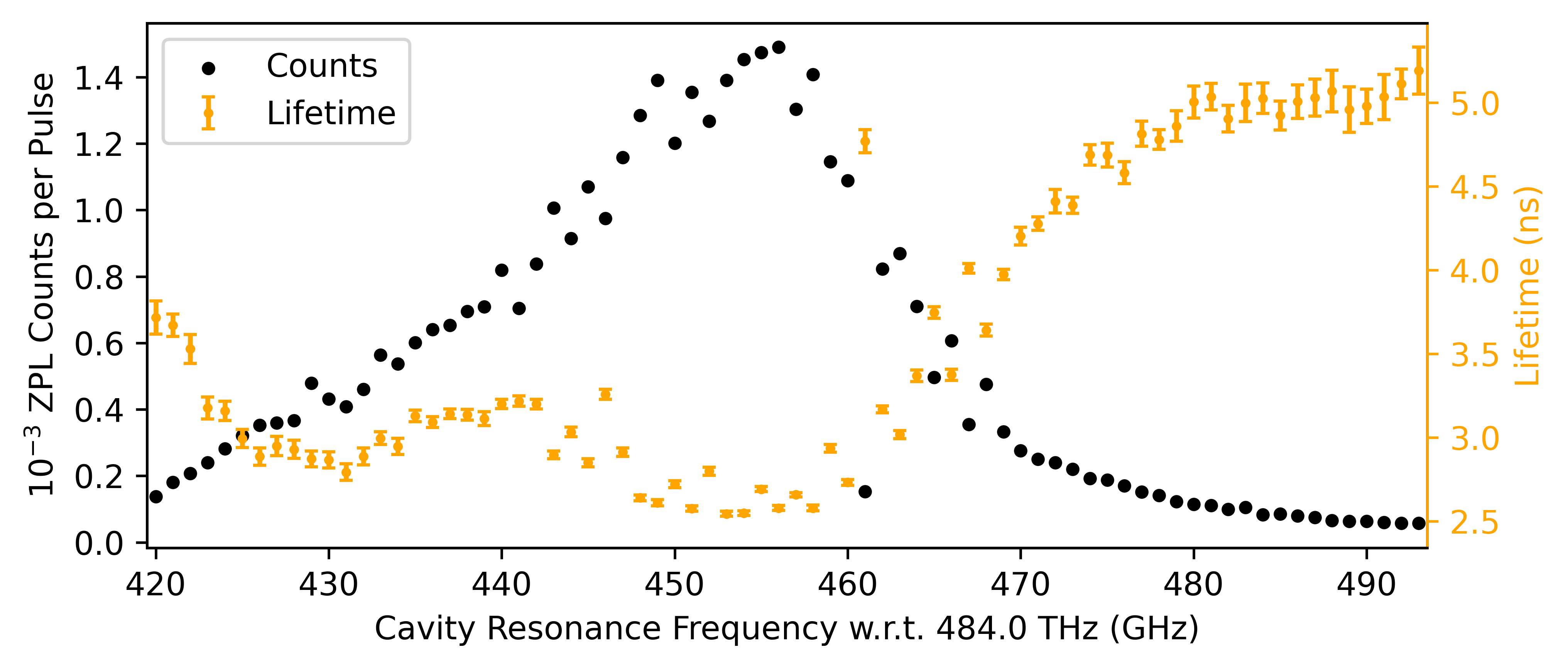}
    \caption{Full dataset of the cavity detuning measurement presented in Fig. \ref{fig:lifetime} (a). The cavity frequency, which is matched to a frequency-stabilized reference laser, is stated in absolute values. Note that in the lifetime measurement of frequency $\unit[461]{GHz}$, the cavity is not set to the correct resonance frequency, which leads to a wrong lifetime value. The error bars represent one standard deviation confidence intervals of the fit.}
    \label{fig:full_detuning_sweep}
\end{figure}

After finding a SnV center with a good coupling strength to the cavity, an etalon filter (full width at half maximum $\approx \unit[45]{GHz}$) is added to the ZPL path to further reduce the background and the polarization detection is optimized on the emitter counts under pulsed, off-resonant $\unit[532]{nm}$ excitation. The coupling is then quantified via cavity detuning dependent lifetime measurements as reported in Fig. \ref{fig:lifetime} (a) and (b). Figure \ref{fig:full_detuning_sweep} depicts the full dataset, that is used in Fig. \ref{fig:lifetime} (a). The investigated emission line is located at around $\unit[455]{GHz}$ (with respect to $\unit[484]{THz}$), whereas at around $\unit[430]{GHz}$ another weaker coupled emission line is present. We choose to analyze the data for frequencies $\unit[\ge455]{GHz}$ only, where leakage of the $\unit[430]{GHz}$ emission line is very small. Note that these measurements under pulsed, off-resonant $\unit[532]{nm}$ excitation were performed last and caused the emitter frequency to jump from the initial $\unit[484.558]{THz}$ to $\unit[484.455]{THz}$. We attribute this jump to the extensive irradiation of the diamond device with high power ($\approx \unit[5]{mW}$) pulsed $\unit[532]{nm}$ light.

\begin{figure}[ht]
    \centering
    \includegraphics[width=\linewidth]{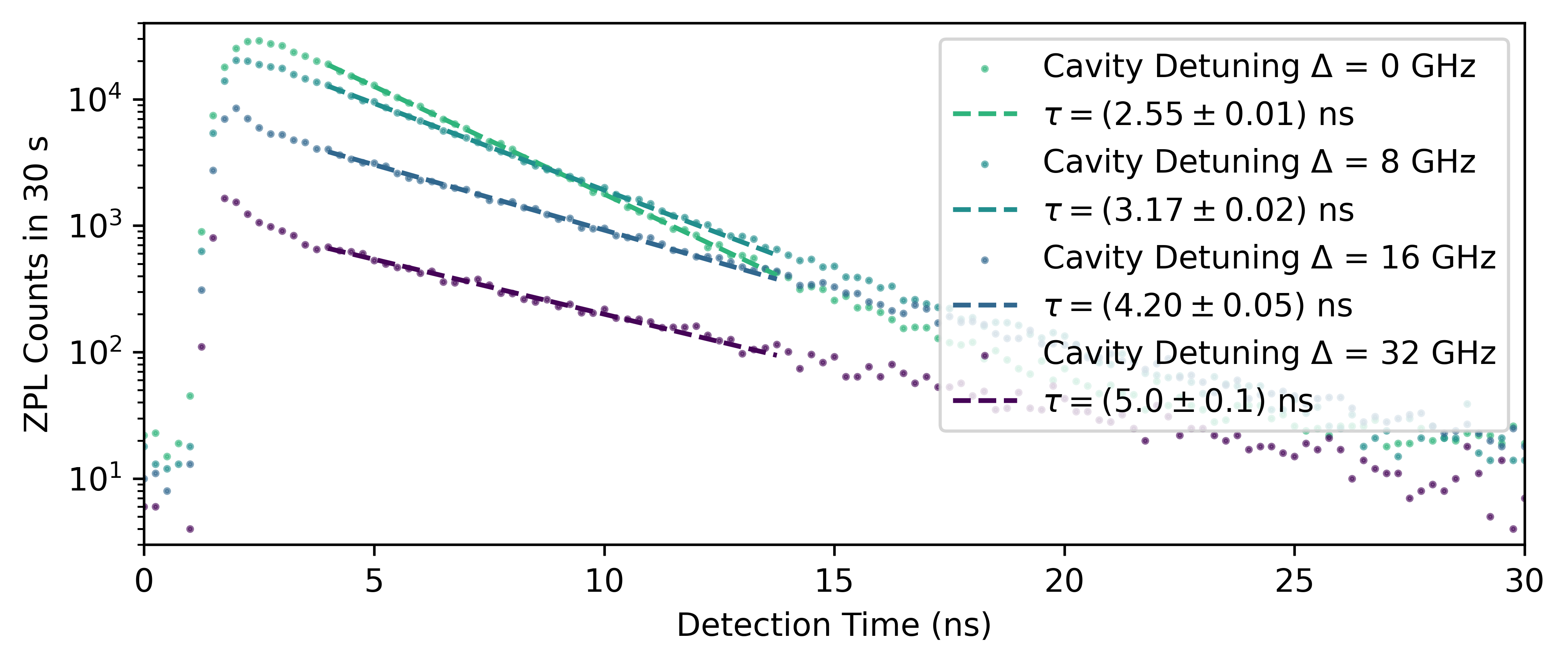}
    \caption{Lifetime measurements and fits (dashed lines) of the data presented in Fig. \ref{fig:lifetime} (b) without normalization. The signal of the SnV center is reduced for larger cavity detuning. Here, an additional fast-decaying signal can be observed for short detection times. We attribute this to background fluorescence, which could stem from the cavity fiber, the mirror coating or the output mirror substrate. The fit window is chosen to exclude this fast decay. A binning of $\unit[0.25]{ns}$ is used.}
    \label{fig:time_traces}
\end{figure}

\begin{figure}[ht]
    \centering
    \includegraphics[width=\linewidth]{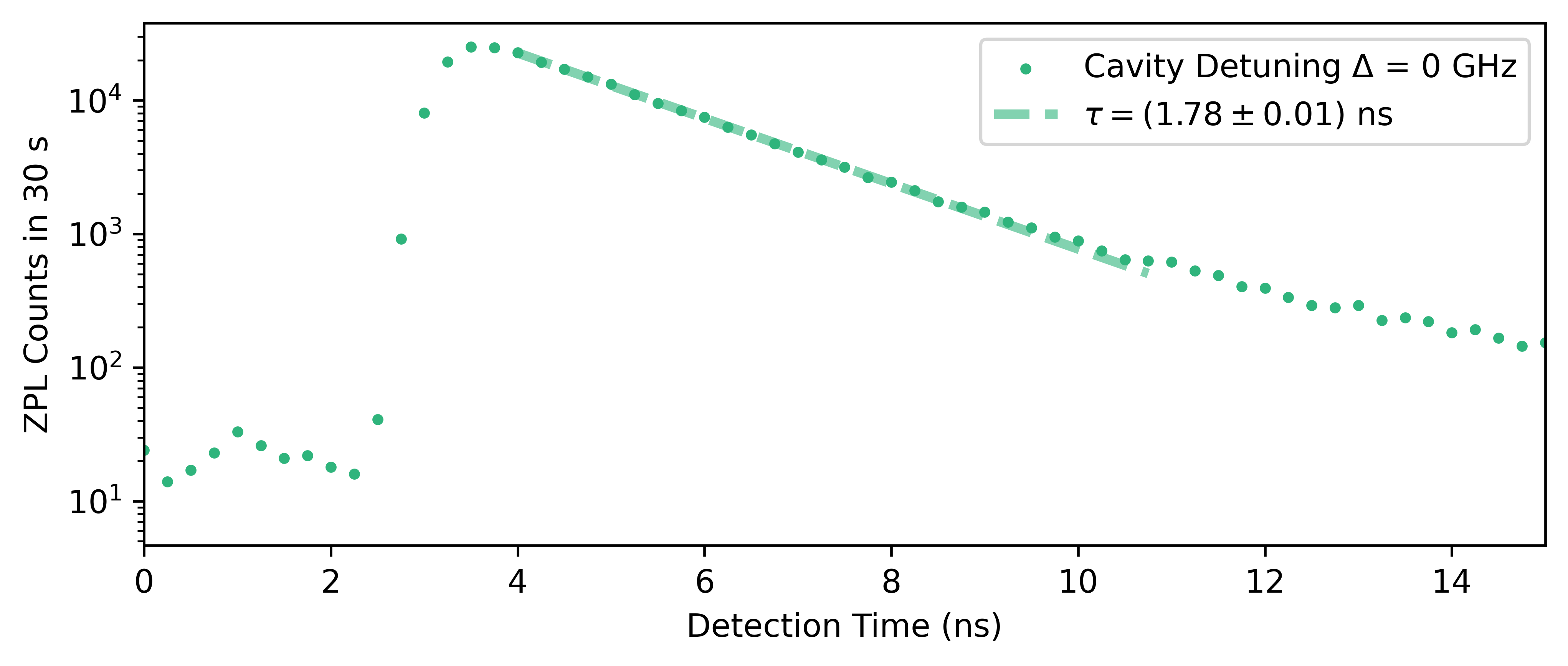}
    \caption{Lifetime measurement of the SnV center, that shows the largest lifetime reduction. This is measured at a different lateral position on the diamond device with a better cavity quality factor of about $8 \times 10^4$ and a comparable mode volume. Note that, in this measurement, the ZPL signal is filtered in a 10 nm window. The change in the exponential slope for longer detection times could stem from weaker coupled SnV centers or not filtered slow-decaying background fluorescence. A binning of $\unit[0.25]{ns}$ is used.}
    \label{fig:best_snv}
\end{figure}

The lifetime measurements of Fig. \ref{fig:lifetime} (b) are shown without normalization in Fig. \ref{fig:time_traces} together with the monoexponential fits, that are used to extract the lifetime.\\
The lifetime measurement of a different SnV center with larger lifetime reduction is shown in Fig. \ref{fig:best_snv}. Note that this is measured at a different lateral cavity position with a better cavity quality factor.

\section{PLE Measurements of SnV Centers with the Microcavity\label{apx:PLE}}
The investigated SnV center shows spectral diffusion and emitter ionization. In the PLE measurements, a conditional off-resonant $\unit[515]{nm}$ laser pulse is applied before scanning the resonant laser over the emission line. This sequence measures the emitter linewidth without contributions of spectral diffusion and the SnV center is likely found in its negative charge state. The usage of the $\unit[515]{nm}$ pulse is conditioned on the detected emitter counts of the previous scan. If a specified peak contrast is not reached the emitter likely ionized and a $\unit[10]{\upmu W}$ $\unit[515]{nm}$ laser pulse is applied for $\unit[50]{ms}$ to repump the charge state. The resonant laser is scanned with about $\unit[180]{MHz/s}$, while we integrate the emitter counts for $\unit[50]{ms}$ at each frequency step. Phonon sideband detection is utilized to filter out residual resonant excitation laser after the microcavity. The microcavity itself strongly filters the PSB emission of the emitter, resulting in a low PSB photon collection rate. Therefore individual PLE scans are postselected and averaged to get a good estimate of the emitter linewidth.

\subsection{Cavity On Resonance PLE Measurement Analysis}
For the cavity on resonance PLE scans a low excitation power corresponding to $\unit[260]{fW}$ is used. This enables the acquisition of the ZPL counts next to the PSB counts with single-photon detectors in the detection. These ZPL count traces are equivalent to transmission dip measurements, which show a significantly better signal to noise than the PSB count traces. We use the transmission dips for postselecting PLE measurements. First, we only consider measurements that exhibit a transmission dip and where the cavity mode is resonant with the laser scan range. Next, each transmission dip is fitted with a Lorentzian function, and further postselection is performed on the resulting fit parameters. The transmission dip, and therewith the emission frequency, has to be in the range $\unit[(558 - 559)]{GHz}$ in the center of the laser scan. In addition, a Lorentzian width $> \unit[100]{MHz}$ and a dip contrast $> \unit[50]{\%}$ is required, which sorts out scans where emitter ionization occurred. The selected PLE scans are centered using the Lorentzian fit center frequencies and finally averaged. For the data presented in Fig. \ref{fig:lifetime} (c), where the cavity is on resonance with the emitter, the average is performed with $\unit[170]{}$ individual scans.

\subsection{Cavity Off Resonance PLE Measurement Analysis}
For cavity off resonance PLE scans the propagation of the scanned resonant laser light into the cavity is highly suppressed. To achieve similar PSB emitter count rates the excitation laser power is increased by a factor of $400$.
In postselection, we use the knowledge if a conditional repump is applied to select scans where no ionization occurred and additionally require a certain emission peak contrast. Then, each PLE scan is fitted with a Lorentzian peak and further postselection is performed based on the resulting fitting parameters. A Lorentzian center frequency in the range of $\unit[(558 - 559)]{GHz}$ and the linewidth in the range of $\unit[(30 - 500)]{MHz}$ is required. Moreover, we select on the Lorentzian peak amplitude as a criterion for a successful fit. The selected PLE scans are centered using the Lorentzian fit center frequencies and finally averaged. For the data presented in Fig. \ref{fig:lifetime} (c), where the cavity is off resonance with the emitter, the average is performed with $\unit[307]{}$ individual scans.

\section{Cavity Transmission Dip Measurements\label{apx:dip}}
\begin{figure}[ht]
    \centering
    \includegraphics[width=\linewidth]{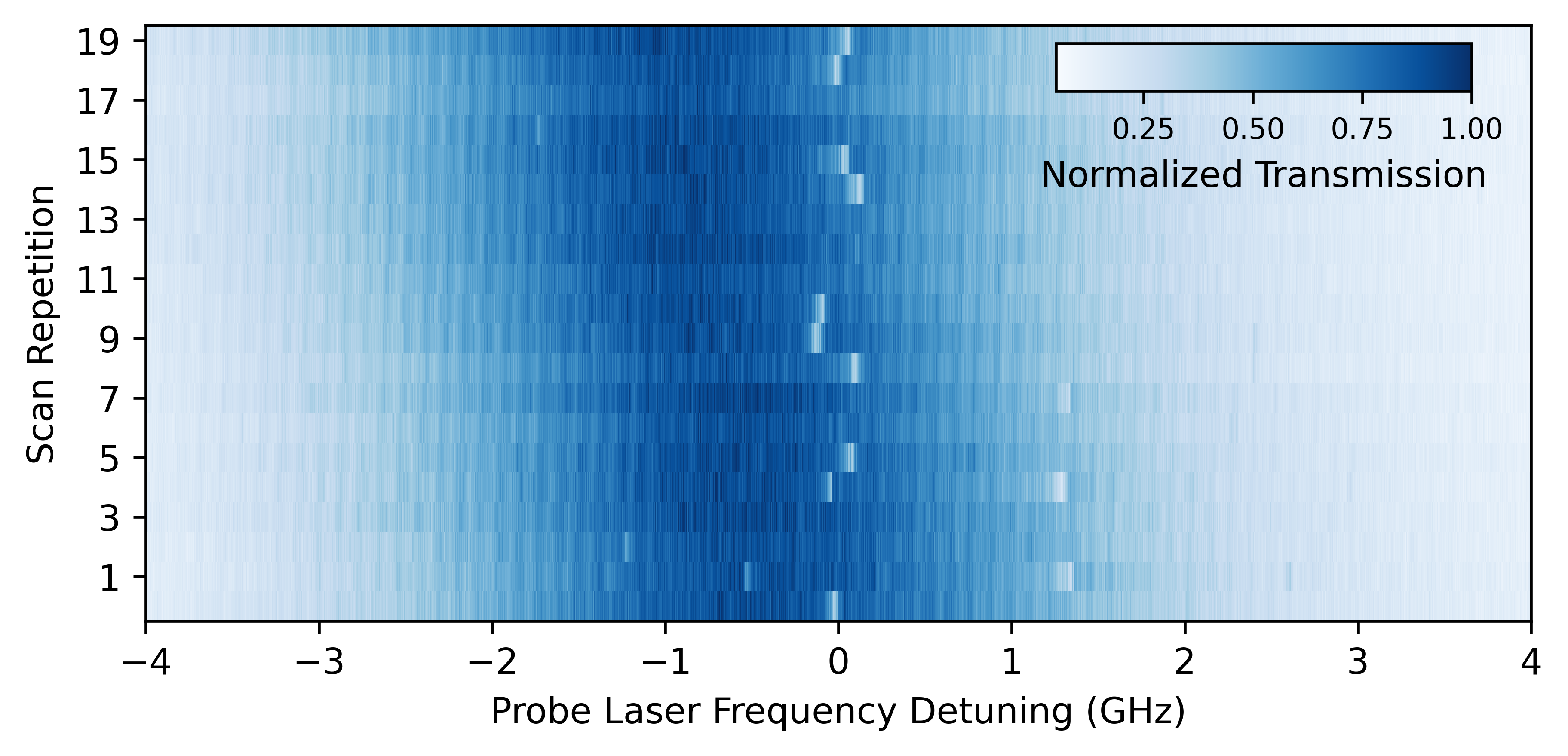}
    \caption{Successively acquired cavity transmission dip measurements. The cavity resonance is not actively controlled during these measurements.}
    \label{fig:transmission_dip_scans}
\end{figure}

The cavity transmission dip measurements are performed by scanning a weak coherent laser over the cavity resonance and detecting the transmitted laser signal with a single-photon detector in the ZPL path. These transmission scans exhibit a good signal to noise ratio allowing to scan the laser faster than in the PLE measurements, which is beneficial to lower the chance of emitter ionization. The measurements depicted in Fig. \ref{fig:dips} (a) and (b) were conducted with a laser power of $\unit[260]{fW}$, a laser scan speed of about $\unit[350]{MHz/s}$ and an integration time of $\unit[20]{ms}$ for each frequency step. A $\unit[50]{\upmu W}$ off-resonant $\unit[515]{nm}$ laser pulse is applied for $\unit[50]{ms}$ before each scan. Figure \ref{fig:transmission_dip_scans} depicts successively acquired cavity transmission dip measurements. These reveal the effect of spectral diffusion of the investigated SnV center. In individual measurements, the SnV center exhibits its narrow emitter linewidth, whereas between measurements spectral diffusion on the order of GHz is observed. We attribute these frequency jumps to changes in the SnV center environment due to the application of the off-resonant $\unit[515]{nm}$ laser before each measurement. Without the application of the off-resonant $\unit[515]{nm}$ laser the frequency jumps between successive measurements are on the order of the emitter linewidth.

\section{Transmission Dip Contrast Measurements\label{apx:contrast}}
\begin{figure}[ht]
    \centering
    \includegraphics[width=\linewidth]{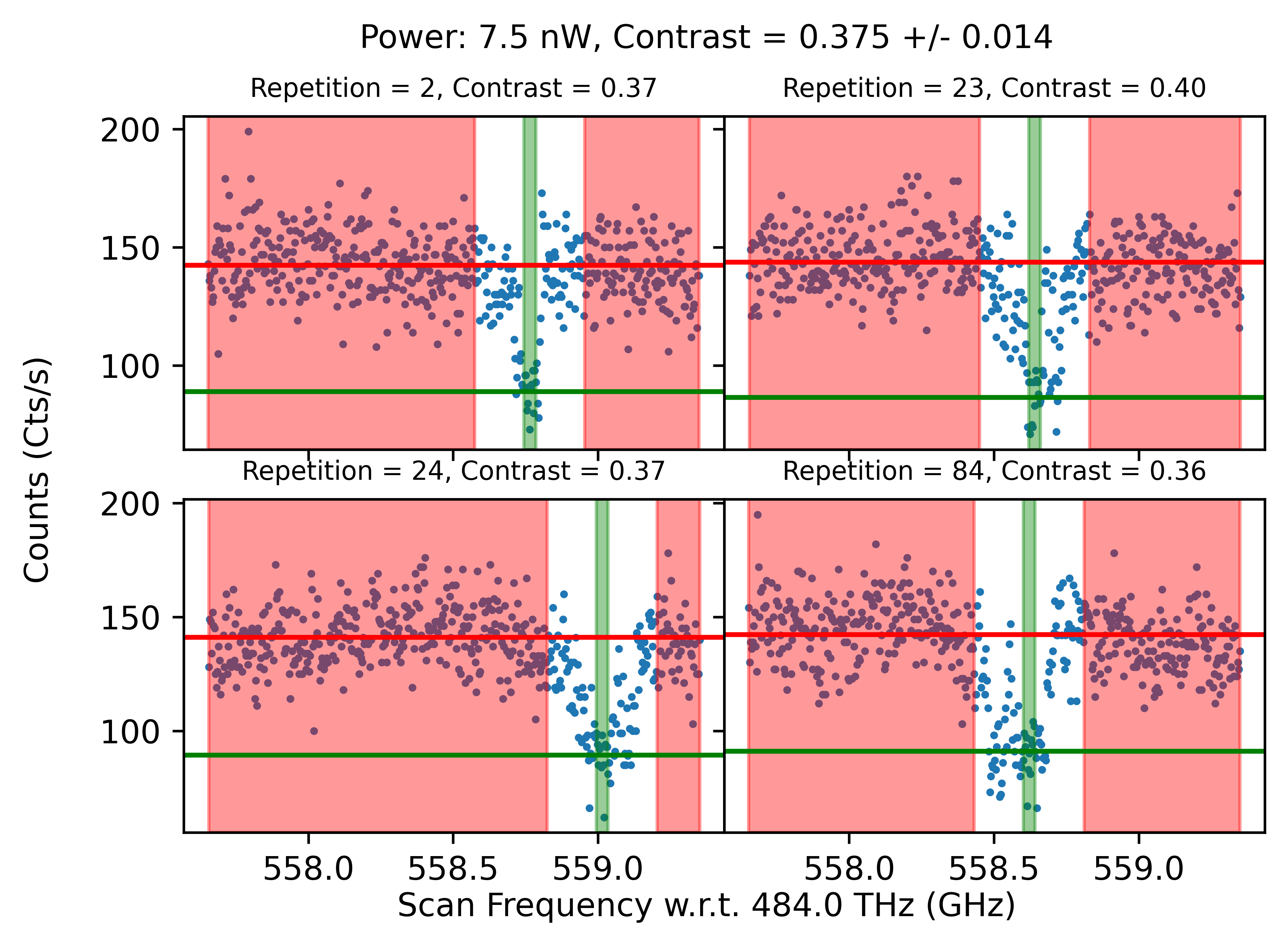}
    \caption{Exemplary analysis of the cavity transmission dip contrasts shown in Fig. \ref{fig:dips} (c) for a set laser power of $\unit[7.5]{nW}$, corresponding to about $\unit[0.1]{}$ photons per lifetime. Four individual scans of the resonant cavity transmission measurements, where a clear dip is visible, are selected. We use a running average of the count rate to automatically identify the position of the dip. 12 data points inside the dip are taken as an average to define the dip depth (green area). Excluding 50 data points to both sides of the dip, the remaining points are averaged to calculated the cavity transmission peak height (red area).}
    \label{fig:dip_contrast}
\end{figure}

The transmission dip contrasts depending on photons per Purcell-reduced excited state lifetime are determined by performing transmission dip measurements with different laser powers. In all of these measurements we use a scan range of $\unit[1.8]{GHz}$ and apply an off-resonant $\unit[515]{nm}$ laser pulse before each scan. The laser scan speed and the frequency step integration time are adjusted for each laser power to acquire transmission dips before they vanish due to emitter ionization. An exemplary transmission dip contrast analysis is shown in Fig. \ref{fig:dip_contrast}. For each power value, we analyze four scans and take the mean contrast value with standard deviation as the dip contrast displayed in Fig. \ref{fig:dips} (c).\\
The experimental data of Fig. \ref{fig:dips} (c) is fit to the simulations by scaling the used excitation powers to the simulated photons per lifetime, which maps the experimentally set laser power to the actual laser power leaving the cavity. A set laser power of $\unit[1.0]{nW}$ measured in our excitation setup corresponds to $\unit[0.015]{}$ photons per lifetime and a power of $\unit[2.6]{pW}$ after the cavity. This power ratio originates from an initial free space to fiber coupling efficiency, fiber and fiber splice losses, fiber mode to cavity mode matching efficiency, and cavity transmission. The cavity transmission is calculated to be \unit[2.4]{\%} and an achievable fiber mode to cavity mode efficiency of \unit[47]{\%} is calculated for the used cavity geometry. The determined power ratio is used to state the resonant laser power values. By comparing the detector count rate with the expected photons per lifetime after the cavity we estimate a cavity mode collection efficiency of about $\unit[12]{\%}$. An independent measurement of the collection efficiency confirms this value.

\section{Photon Statistics Measurement of the Cavity Transmission on Emitter Resonance\label{apx:statistics}}
The measurement, depicted in Fig. \ref{fig:dips} (d), shows the photon statistics of the light, that is transmitted through the cavity on emitter resonance. The transmitted light is collected by a single-mode fiber in the ZPL path and guided by a 50:50 fiber beam splitter to two single-photon detectors, which are connected to time-tagging electronics. In this measurement a probe laser power of $\unit[2.6]{pW}$, a scan speed of about $\unit[1.8]{GHz/s}$ and a frequency step integration time of $\unit[5]{ms}$ is used. To exclusively acquire data on emitter resonance we only trigger the time tagging electronics to record for the next $\unit[5]{ms}$ if the transmission dip contrast is $\unit[>35]{\%}$ in the previous frequency step. Before each scan a $\unit[10]{\upmu W}$ off-resonant $\unit[515]{nm}$ laser pulse is applied for $\unit[50]{ms}$. In the simulated graph of Fig. \ref{fig:dips} (d) the detunings between emitter and probe laser due to the transmission dip trigger threshold of $\unit[35]{\%}$ are taken into account. Emitter-cavity detunings due to spectral diffusion and small cavity drift during the course of the experiment are not captured by the measurement routine nor by the simulations.

\section{Summary of System Parameters\label{apx:parameters}}
An overview of measured, estimated and simulated values of this work are summarized in Table \ref{tab:summary}.

\begin{table}[b]
\caption{\label{tab:summary}}
\footnotesize
\begin{ruledtabular}
\begin{tabular}{ll}
\textrm{Parameter}&
\textrm{Value}\\
\colrule
Input mirror radius of curvature & $\unit[15.7]{\upmu m}$ \\
Cavity air gap & $\unit[6.50]{\upmu m}$ \\
Diamond thickness & $\unit[3.72]{\upmu m}$ \\
Hybrid cavity mode number $q$ & $\unit[50]{}$ \\
Effective cavity length $L_\text{eff}$ \cite{van_dam_optimal_2018} & $\unit[10.8]{\upmu m}$ \\
Estimated cavity beam waist $\omega_0$ \cite{van_dam_optimal_2018} & $\unit[1.24] {\upmu m}$\\
Estimated cavity mode volume $V$ \cite{van_dam_optimal_2018} & $\unit[55]{\lambda^3}$\\
Cavity Lorentzian linewidth $\kappa/2\pi$ &  $\unit[(6.86\pm0.05)]{GHz}$ \\
Root mean square cavity length fluctuations & $\unit[27]{pm}$ \\
Cavity mode dispersion slope & $\unit[46]{MHz/pm}$ \\
Cavity quality factor $Q$ & $7 \times 10^4$ \\
Cavity finesse $\finesse$  &  $\unit[830]{}$ \\
Calculated Purcell factor $F_P$ \cite{van_dam_optimal_2018}& $\unit[6.9]{}$\\ 
Calculated cavity transmission & $\unit[2.4]{\%}$\\ 
SnV center natural lifetime $\tau$ & $\unit[(5.0 \pm 0.1)]{ns}$ \\
%Purcell-reduced SnV center lifetime $\tau_P$ & $\unit[(2.55 \pm 0.01)]{ns}$ \\
SnV center lifetime-limited linewidth $\gamma/2\pi$ & $\unit[32]{MHz}$ \\
%Cooperativity $C$, measured by lifetime & $\unit[0.96 \pm 0.05]{}$ \\ 
%Estimated Purcell factor (from lifetime) &  \unit[3.8]{}\\ 
SnV center linewidth $\gamma^\prime/2\pi$ & $\unit[(77.6 \pm 0.8)]{MHz}$ \\ 
Purcell-enhanced emitter linewidth $\gamma^\prime_P/2\pi$ & $\unit[(126 \pm 4)]{MHz}$ \\ 
%Cooperativity $C$ (from linewidth) & $\unit[1.5 \pm 0.2]{}$ \\
Vibration-corrected cooperativity $C$ & $\unit[1.7 \pm 0.2]{}$\\
Vibration-corrected coherent cooperativity $C_\text{coh}$& $\unit[0.69 \pm 0.07]{}$ \\ 
Single photon Rabi frequency $g$ & $\unit[(300\pm20)]{MHz}$ \\
\end{tabular}
\end{ruledtabular}
\end{table}

\clearpage
\bibliography{bib}

\end{document}